# A Comprehensive Survey on Spectrum Sharing Techniques for 5G/B5G Intelligent Wireless Networks: Opportunities, Challenges and Future Research Directions

Anita Patil[1], Sridhar Iyer[2], Onel L. A. López[3], Rahul J Pandya[4], Krishna Pai[5], Anshuman Kalla[6], and Rakhee Kallimani[7]

[1]Department of ECE, S.G. Balekundri Institute of Technology, Belagavi, KA, India- 590010.
Email: anitap@sgbit.edu.in

[2]Department of ECE, KLE Technological University Dr. MSSCET, Belagavi, KA, India- 590008.
Email: sridhariyer1983@klescet.ac.in

[3]Faculty of Information Technology and Electrical Engineering, University of Oulu, Finland- 90014.
Email: onel.alcarazlopez@oulu.fi

[4]Department of Electrical Engineering, Indian Institute of Technology-Dharwad, KA, India- 580011.
Email: rpandya@iitdh.ac.in

[5]Department of ECE, KLE Technological University Dr. MSSCET, Belagavi, KA, India- 590008.
Email: krishnapai271999@gmail.com

[6]CGPIT, Uka Tarsadia University (UTU), Gujarat, India - 394350
Email: anshuman.kalla@ieee.org and anshuman.kalla@utu.ac.in

[7]Department of EEE, KLE Technological University Dr. MSSCET, Belagavi, KA, India- 590008.
Email: rakhee.kallimani@klescet.ac.in

**The increasing popularity of the Internet of Everything and small-cell devices has enormously accelerated traffic loads. Consequently, increased bandwidth and high data rate requirements stimulate the operation at the millimeter wave and the Tera-Hertz spectrum bands in the fifth generation (5G) and beyond 5G (B5G) wireless networks. Furthermore, efficient spectrum allocation, maximizing the spectrum utilization, achieving efficient spectrum sharing (SS), and managing the spectrum to enhance the system performance remain challenging. To this end, recent studies have implemented artificial intelligence and machine learning techniques, enabling intelligent and efficient spectrum leveraging. However, despite many recent research advances focused on maximizing utilization of the spectrum bands, achieving efficient sharing, allocation, and management of the enormous available spectrum remains challenging. Therefore, the current article acquaints a comprehensive survey on intelligent SS methodologies for 5G and B5G wireless networks, considering the applications of artificial intelligence for efficient SS. Specifically, a thorough overview of SS methodologies is conferred, following which the various spectrum utilization opportunities arising from the existing SS methodologies in intelligent wireless networks are discussed. Subsequently, to highlight critical limitations of the existing methodologies, recent literature on existing SS methodologies is reviewed in detail, classifying them based on the implemented technology, i.e., cognitive radio, machine learning, blockchain, and multiple other techniques. Moreover, the related SS techniques are reviewed to highlight significant challenges in the B5G intelligent wireless network. Finally, to provide an insight into the prospective research avenues, the article is concluded by presenting several potential research directions and proposed solutions.**

## I. Introduction

THE fifth generation (5G) wireless networks have been commercialized and are being deployed globally, resulting in fierce competition between mobile network operators and vendors to attain market primacy. The research on 5G wireless networks commenced with the aim of enabling innovative services across the different industry sectors, and envisioned a major technological shift from the 4G/Long Term Evolution (LTE) concerning total spectral-efficiency (SE) and energy-efficiency (EE), in addition to latency and reliability. [1] [2]. Further, in comparison to the 4G/LTE networks, which split the radio access architecture into two elements, viz., the radio head which is remote and the unit of baseband, the 5G wireless networks introduce a three-element split architecture which includes centralized unit (CU), distributed unit (DU), and radio unit (RU) [3]. Hence, a network designer can decide

Corresponding author: Sridhar Iyer (email: sridhariyer1983@klescet.ac.in).

the network functionality placement within these building blocks to attain the desired network requirements, including delay and network throughput. Specifically, the Ethernet-enabled F1 interface protocol stack will cover how the CU and the DU will communicate. Further, the functionalities provided by CU, DU, and RU can be included within a single device. Hence, based on the network architecture adopted by a network operator, using either of the interfaces, the cell sites can be connected to the remaining infrastructure [4].

Towards the end of 2017, 3rd Generation Partnership Project (3GPP) standards body approved the intermediate specifications set for the 5G wireless networks, which shifted the focus on provisioning enhanced mobile broadband (eMBB) feature. Simultaneously, 5G new-radio (5G NR) was defined which enables the 5G wireless networks to furnish enhanced bandwidth and lower latency [8]. Therefore, the 5G standards comprise advancements from the existing LTE network in the form of the 5G NR technologies. The International Telecommunication Union (ITU) has hence identified numerous 5G application services including Massive Machine Type Communication (mMTC), eMBB, and Ultra-Reliable Low-Latency Communication (URLLC) [2]

Further, 5G non-standalone (NSA) architecture was also defined, which permits the 5G NR to employ the 5G wireless network for the downlink (DnL) operation simultaneously, enabling the 4G/LTE network users to communicate via the uplink (UpL). This implies that the core 5G NSA service is built over the existing 4G/LTE network. Also, 5G NSA enables features such as, uRLLC and mMTC applications and supports a complete core operation wherein, the 5G network can operate on both the UpL and the DnL, in addition to provisioning further improvements on latency and device connectivity [9] [10]. Thereafter, the 5G standalone (SA) was defined to connect the 5G radio directly with the core 5G network. This results in the control signaling being independent of the existing 4G/LTE networks i.e., the 5G services operate independently without interacting with the 4G/LTE networks. With the 5G wireless network implementation, each user will experience the highest and the lowest data rate of 20 Gbps and 100 Mbps, respectively [11]. Residential customers and enterprises will be benefited from this broadband wireless network using the pre-5G or the 5G technology(s) for access such as massive Multiple Input Multiple Output (M-MIMO), full dimensional MIMO (FD-MIMO), and millimeter-wave (mm-wave) technology(s) for radio access. According to the vision of ITU, a minimum of 800 MHz spectrum is needed to provision large capacity 5G networks such as, Fixed Wireless Access and hotspots [5].

However, after almost a decade of intense academic and industrial research on the 5G wireless networks and the subsequent commercial deployment, it has become clear that the 5G wireless networks will fall short in supporting the vision of the Internet of Everything (IoE), which will aim to provision the advanced bandwidth hungry applications such as augmented-reality (AR), virtual-reality (VR), mixed-reality (XR), tele-presence, and Industry 5.0 requiring either least delay, connections with reliability or accessing of the Internet via the mm-wave/Tera-Hertz (THz) frequency [6].

Therefore, research community has commenced intensive research on the successor technology, the Beyond 5G (B5G) or the 6G wireless networks [7]. The B5G network is envisioned to overcome shortcomings of 5G wireless networks by adopting a cell-free architecture and operating in sub-6 GHz, mm-wave, or THz spectrum. However, transforming such speculative ideas into real-time commercial deployments is a big challenge. Fig. 1 shows the technical specifications of the 4G, 5G, and B5G wireless networks.

### A. Article Motivation

In recent years, advanced applications such as connected robotics, autonomous systems, smart healthcare, intelligent transportation, cloud-based services, AR, and VR have become an integral part of the intelligent human society's lifestyle. [2]. Wireless networks support these and future applications, which are expected to demand connectivity for around 50 billion devices by 2025 as part of Internet of Things (IoT) evolution [12]. In general, devices such as street lighting, electronic appliances, sensors, actuators, robots, and vehicles will seamlessly connect device-to-device (D2D) and to the internet (i.e., mMTC) [2]. This requires enormous amounts of bandwidth which can be provided by operating at the mm-wave and THz frequency bands [13]. However, in some cases, multiple issues such as shadowing, scattering, high path-loss, and interference may still tilt the balance in favor of the sub-6 GHz spectrum utilization. All in all, there is a fundamental need of intelligent spectrum sharing (SS) strategies, which include the following four critical steps [14]:

1) **Spectrum allocation** - comprises techniques and rules for efficiently allocating spectrum to secondary users. Secondary users must utilize the spectrum opportunistically simultaneously avoiding interference to primary users who have prime operation rights of the spectrum.
2) **Spectrum access** - comprises coordination mechanisms among many secondary users aiming to access spectrum. They ensure conflict avoidance between primary users and secondary users.
3) **Spectrum sensing** - used for detecting whether an incumbent or a primary user appears and for determining the status/availability of the spectrum slot.
4) **Spectrum hand-off** - manages the spectrum switching/re-allocation when (i) secondary users currently using a spectrum slot cause collision to a primary user, and this requires the switching of secondary user to alternate spectrum slot, (ii) geography of the primary user remains the same; however, there occurs a change in geography of the secondary user, which requires the latter to switch to other appropriate spectrum bands (slots), and (iii) spectrum bands (slots) utilized by the secondary users are inadequate for meeting desired needs; hence, secondary users must switch to more favorable spectrum bands (slots).

In addition, the issue of carrier aggregation (CA) must be addressed for designing an optimal SS framework, enabling the allocation of multiple spectrum resources efficiently amongst the users [22]. A detailed channel assessment also needs to be conducted as response to UpL grant for determining if unlicensed spectrum is available [23]. To overcome spectrum allocation and management-related issues, multiple machine learning (ML) enabled spectrum allocation methods have been implemented in the existing studies [24]. However, control of wireless network spectrum is not dynamic and this results in spectrum loads which are unbalanced thereby resulting in capacity related issues. Hence, it is required to formulate solutions which follow the extended dynamic spectrum access method for provisioning balance of load with quality within spectrum available [25]. Further, effective management of spectrum is also essential to obtain maximum useage of the resource with desired Quality of Service (QoS) maximization.

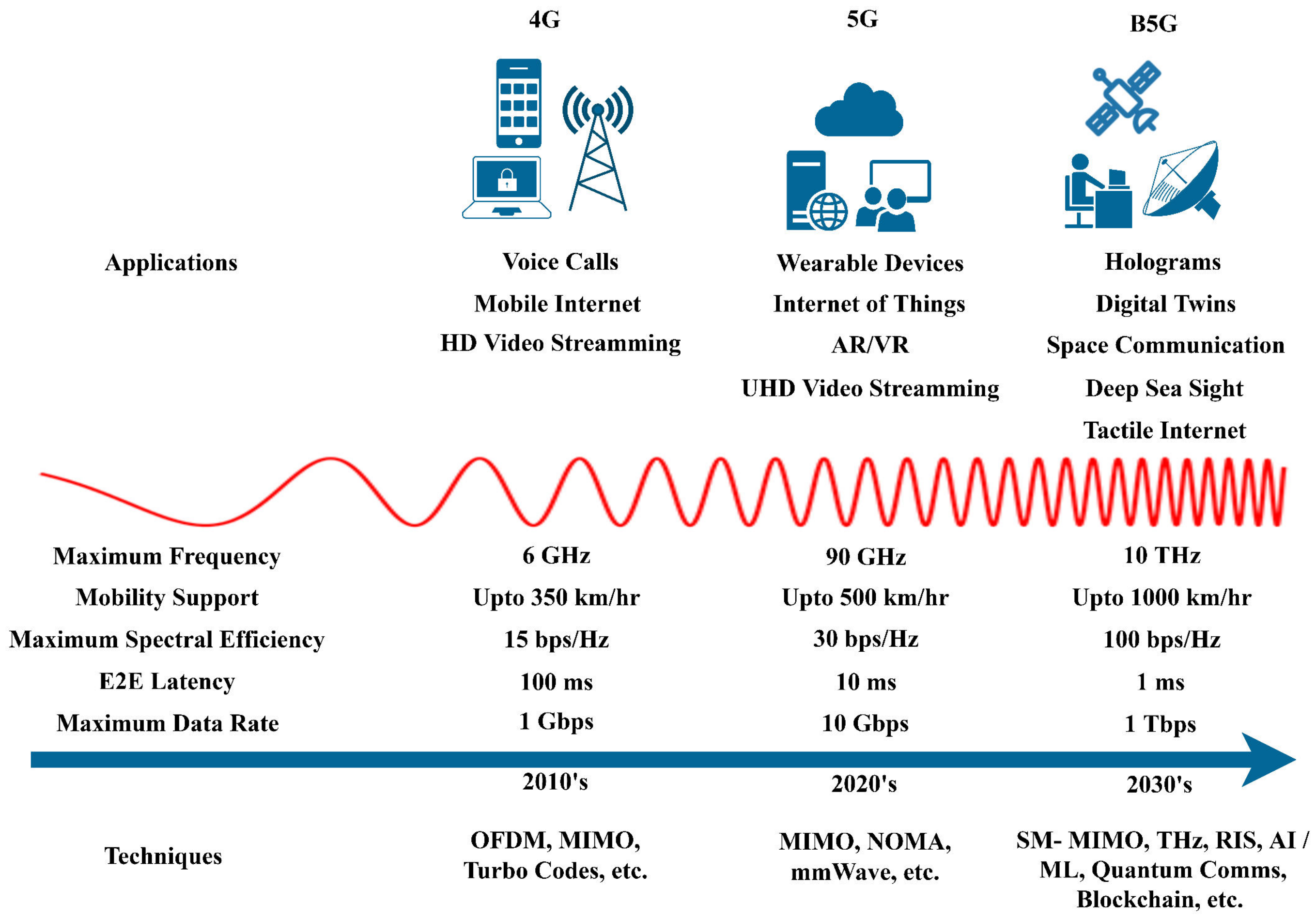

Fig. 1. Technical Specifications of 4G, 5G and B5G wireless networks [5], [6], [7].

Many studies in the literature have also implemented cognitive radio (CR) enabled techniques such as matched filter and cyclo-stationary feature detection, to solve SS issues [26], [27], [28]. Further, distributed cooperative sensing has been used owing to which the secondary users can receive information from the neighbors, thereby enabling them to individually improve the rate of detection of a primary user signal [29]. Artificial intelligence (AI)/ML techniques have also been used recently with CR to improve network SE and EE, and for managing spectrum [30], [31]. However, extensive research is required on the issues related to using various parameters for optimization parameters considering varied scenarios and ensuring processing in real time.

The B5G networks will implement small cell/cell free architectures since such networks will exploit high frequency bands for serving numerous users [32]. However, current spectrum allocation methods will be ineffective in view of delivering the optimum efficiency within the small cell or cell-free network [33]. Hence, efficient algorithms need to be formulated for classifying all small cells in accordance with locations as per geography and radius of interference. Also, considering limited network backhaul capacity, efficient spectrum allocation methods are required for performance improvement considering increased throughput, reduced latency, and savings in power for small cell networks, which may require leveraging complementary technology(s) such as Wi-Fi and LTE [34], [35]. Further, efficient design of the advanced AI/ML-enabled clustering approaches is necessary for managing the spectrum, simultaneously ensuring enhancement of small cell's throughput and efficiency, and improving the system coverage [36].

Limited coverage remains a key issue when operating in the mm-wave and the THz spectrum. Indeed, accurate coverage analysis may rely on detailed stochastic geometric

TABLE I
SUMMARY OF RECENT SURVEYS RELATED TO SS IN 5G, AND B5G.

| Reference | Key Contribution | Limitations w.r.t our survey |
|---|---|---|
| [15] | • Survey of various SS techniques such as trading, relaying, and routing is conducted.<br>• A comparison of the spectrum usage for various wireless generations is presented.<br>• 5G spectrum requirements and techniques for spectrum management are listed. | • Criticality in regard to the integration of different bands is not addressed since the article does not highlight the solutions for issues related to interference management and small-cell deployment. |
| [2] | • A detailed list of the essential 5G-enabling technologies is provided followed by a brief overview of every technology.<br>• Various SS techniques are classified.<br>• An extensive review on SS techniques in regard to 5G networks is discussed. The outcome of study details the network architecture, spectrum allocation and different methods to access spectrum.<br>• Related studies are summarised, compiled, and presented as a table.<br>• The issues and challenges in current SS implementations and potential improvements to support advances in 5G are discussed.<br>• Realization of interoperability, context awareness, learning ability, self-optimization, dynamic spectrum management, adaptive decoding, and self-healing by CR technique is presented.<br>• It is specified that cognitive dynamic network architecture uses cognitive methods for exploiting underutilized spectrum in a flexible and intelligent manner. | • Major issue in dense networks viz., multi-user interference, especially in licensed bands, is not discussed.<br>• Interference management, required for the CR technique to handle inter-network interference, is not highlighted. |
| [16] | • Beam-collision interference scenarios are introduced which exclusively occur in the mm-wave spectrum.<br>• Centralised resource allocation algorithms from 4G/LTE are extended to include beam-scheduling among multiple base stations to avoid excessive interference.<br>• Features covered in 5G mm-wave bands such as mandatory beam-forming and infinite spatial reuse, are highlighted. | • Solutions for spectrum co-existence are not addressed. |
| [17] | • A genetic algorithm design for dynamic SS is presented.<br>• Three operations of selecting, crossing, and mutating in the proposed algorithm include multiple parameters such as crossover probability and mutation probability.<br>• In practical applications, multiple tests are needed to find the optimal parameters; hence, an adaptive genetic algorithm is introduced.<br>• The evaluation function mechanism is adopted in search process to reduce the complexity. | • Networking technology in regard to wireless spectrum resource(s) scarcity is not addressed, and a detailed study of IoT technology is not provided. |
| [18] | • State of art for classical SS techniques is detailed.<br>• Different Operating modes of CR, derived from implementing FD tool in CR are comprehended.<br>• Siginficant role of ML and DL in improving SS is surveyed.<br>• Detailed survey on recent achivements as a service and DSS for ToT/WSN is discussed.<br>• Applications of CR and SS in 5G and B5G is discussed.<br>• Investigated and discussed the recent trends and potential challenges in future wireless communication.<br>• SS in IoT/WSN and latest achievements in spectrum sensing as a service and DSS for IoT/WSN networks are surveyed.<br>• Possible applications of CR, especially SS in 5G and B5G, are discussed.<br>• New trends and challenges in future wireless communication technologies are also discussed and investigated.<br>• Prominent issues addressed are: In the possible scenarios (i) SS is not performed in which secondary user is unaware of primary user during spectrum slot. (ii) secondary throughput is affected due to silence during sensing time rather than silence during sensing slot. | • Homogeneity assumption of primary users is not addressed. |
| [19] | • The system steady state of the model is analysed using the Markov chain, and performance measures are obtained.<br>• To optimize the system socially, the Nash equilibrium and social optimization strategies are investigated.<br>• Pricing policy with an appropriate spectrum admission fee is exhibited. | • Presented pricing policy does not include the results of all user(s) types that arrive to access spectrum. |
| [20] | • A multi-source mechanism for dynamic adjustment of occupied frequency bands is proposed.<br>• Instead of relying only on radio-related information, a system that collects data from various sources is discussed.<br>• The policy-defined dynamic spectrum access is presented.<br>• Proposed method uses fuzzy and soft connections of multiple contextual information sources to limit radio resource use.<br>• Combined data, as a source of information about users, is presented. | • Using contextual data on user density from various sources in the Open Radio Access Network (ORAN)-based networks is possible; however, such an analysis is not presented. |
| [21] | • Details on state of art techniques proposed to cater to multi-cast services in CR wireless networks are presented.<br>• Possible design guidelines that can be adopted in view of the multi-cast services in 5G networks have been detailed.<br>• Key research issues have been identified, and future research directions have been discussed.<br>• The survey details different SS models adopted in CR wireless networks, which are implemented to fulfill the main aim of multi-cast services, viz., end-to-end communication.<br>• The multi-cast traffic nature, fundamental techniques and methodologies which can be implemented to enhance the efficiency of multi-cast services, and strategies to minimize interference experienced by the incumbents, have been presented. | • The survey is presented with a view of CR-enabled SS techniques.<br>• Other related technologies for SS are not discussed. |
| [13] | • The idea of 'expansive networks' is conceptualized.<br>• The security implications of expansive networks are detailed.<br>• Spectrum resource at PHY layer is elaborated as an example of the vital aspect of an expansive network.<br>• Two key enablers, viz., Distributed Ledger Technology (DLT) and network intelligence via ML, are discussed<br>• Potential SS issues and motivation for expansive network in B5G networks are described. | • The survey is presented with a view of expansive networks.<br>• Other related technologies for SS are not discussed. |
| **Our Survey** | • SS techniques relevant to 5G networks are discussed.<br>• In regard to limitations of existing surveys, solutions offered by the next-generation B5G wireless networks have been highlighted.<br>• Relevant solutions for SS issues are discussed.<br>• Recent literature demonstrating the limitations of SS techniques in 5G networks and future technology and solutions provided by B5G networks are surveyed. | - |

including models for (i) channels which are realistic, and (ii) radiations by antenna [37]. Even though channel models from literature present details of mm-wave propagation for cellular scenario, investigation is mandated for capturing fade and propagation considering THz frequency [38]. Methods dependent on AI/ML are needed for providing clustering and efficient spectrum allocation for mm-wave and THz systems and for optimizing high spectrum compressed sensing in 5G/B5G image transmission [39].

Concerning the M-MIMO scenario, low-cost/complex hybrid pre-coding techniques are needed for modeling transmitters which are efficient, and test beds for mitigating jamming [40]. Also, AI/ML techniques can be used for predicting different characteristics of channel and to create dataset framework for beam-forming M-MIMO [41]. Lastly, explainable AI-controlled-based architectures are also required to overcome exiting limitations simultaneously conducting spectrum assignment, optimization of energy, and minimizing interference [42]. Also, use of standardized techniques for cancelling the interference must be investigated [43].

Overall, for the intelligent 5G/B5G wireless networks, it is necessary to address concerns such as efficiently sharing and effectively managing the spectrum in heterogeneous networks for synchronizing transmission at similar frequency. The multiple open research issues motivate to conduct the survey presented in the current article, which stems from the fact that the existing literature has already identified that the features promised within the deployed 5G technology have major shortcomings and require multiple advancements to provision the next-generation intelligent applications and use-cases. In this regard, towards the end of this survey article, we also present critical prospective challenges in research and propose the corresponding directions of research.

### B. Our Contribution

Although an extensive body of literature has discussed the SS techniques, few detail limitations of the existing SS methods and the key aspects to be developed for sharing the spectrum in the future wireless network. Aiming to fill this void, this paper surveys SS techniques for the intelligent 5G/B5G wireless networks. First, the various SS techniques are overviewed and the multiple types of SS methods are discussed. Next, the various spectrum utilization opportunities arising from the existing SS techniques in intelligent wireless networks are detailed. Then, the critical limitations of the existing SS methods are highlighted by conducting a thorough review of the recent literature. This review classifies the corresponding studies based on the implemented technology, which includes CR, ML, blockchain, and others. In addition, the significant advancements which are required in the SS techniques for implementation in the B5G intelligent wireless networks are highlighted. The article is concluded by presenting multiple potential research directions and solutions thus providing a path towards prospective research.

The key contributions of this paper are:

1) **Highlight limitations of existing SS techniques**: This paper aims at exploration and discussion of literature on features and limitations of the existing SS techniques for intelligent wireless networks.
2) **Present SS techniques taxonomy**: Taxonomy of SS techniques in the existing intelligent wireless networks is devised in the paper. This helps to understand the limitations and identify the solutions which must be included within the B5G intelligent wireless networks.
3) **Detail research studies, and summarize standardization approaches**: Existing studies which focus on SS techniques for 5G and B5G intelligent wireless networks are detailed and summarized. The related standardization activities are also summarized in this article.
4) **Propose roadmap for future research directions**: Presenting and discussing the various existing and prospective challenges in research, and the related probable solutions with an aim to enhance the research pertaining to the requirements of wireless networks enabled through intelligence.

A summary of the current survey article's contribution in comparison to recent surveys on SS in 5G and B5G is presented in Table I. The taxonomy of the SS techniques addressed in the current survey article is shown in Fig. 2. Further, the acronyms used commonly throughout the survey article are listed in Table II.

TABLE II
THE LIST OF IMPORTANT ACRONYMS

| Acronym | Definition |
|---|---|
| 4G | Fourth Generation |
| 5G | Fifth Generation |
| 5G NR | Fifth Generation New Radio |
| 5G NSA | Fifth Generation Non-Standalone |
| 5G SA | Fifth Generation Stand Alone |
| AI | Artificial Intelligence |
| B5G | Beyond Fifth Generation |
| CA | Carrier Aggregation |
| CR | Cognitive Radio |
| DnL | Down-link |
| DR | Dynamic Routing |
| DSA | Dynamic Spectrum Assignment |
| DSS | Dynamic Spectrum Sharing |
| EE | Energy-Efficiency |
| eMBB | Enhanced Mobile Broadband |
| eURLLC | Enhanced Ultra Reliable Low-Latency Communication |
| HeTNeT | Heterogeneous Networks |
| IoE | Internet Of Everything |
| IoT | Internet Of Things |
| LBT | Listen Before Talk |
| LTE | Long Term Evolution |
| MDP | Markov Decision Process |
| MIMO | Multiple Input Multiple Output |
| ML | Machine Learning |
| M-MIMO | Massive Multiple Input Multiple Output |
| mMTC | Massive Machine Type Communication |
| mm-Wave | Millimetre Wave |
| NOMA | Non-Orthogonal Multiple Access |
| OFDM | Orthogonal Frequency Division Multiplexing |
| O-RAN | Open-Radio Access Network |
| QoE | Quality Of Experience |
| QoS | Quality Of Service |
| RA | Resource Allocation |
| RL | Reinforcement Learning |
| SE | Spectral Efficiency |
| SNR | Signal to Noise Ratio |
| SS | Spectrum Sharing |
| THz | Terahertz |
| UDN | Ultra-Dense Network |
| UIoT | Ubiquitous Internet Of Things |
| UpL | Up-link |
| URLLC | Ultra-Reliable Low-Latency Communication |

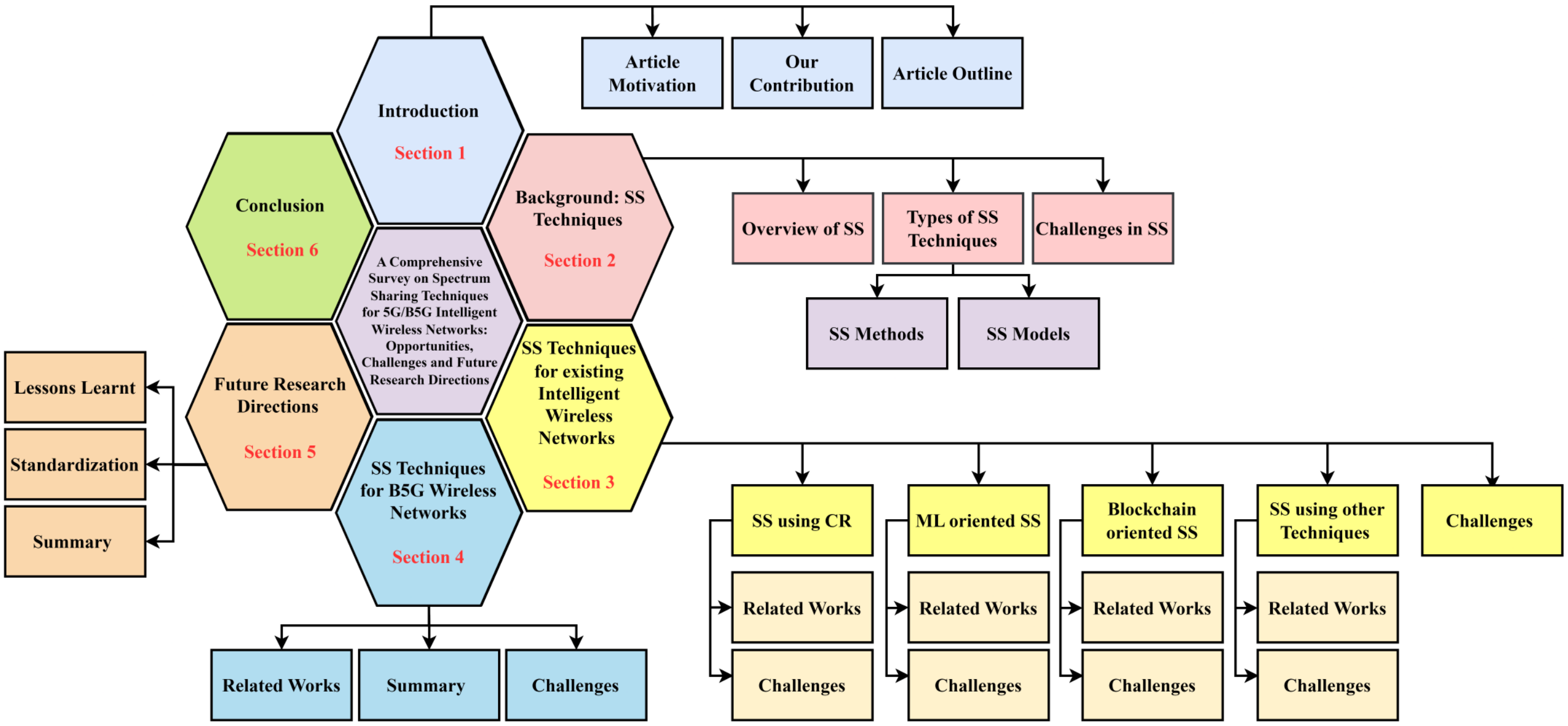

Fig. 2. Taxonomy of the presented SS techniques in the current article.

### C. Article Outline

This paper is organized as follows. In **Section II**, we provide background of SS. In specific, we overview SS techniques, detail the types of SS methods, and present the key SS challenges in intelligent wireless networks. **Section III** reviews the literature on SS techniques in the existing intelligent wireless networks based on the implemented technology, viz., CR, ML, blockchain, and others. In **Section IV**, we highlight the major advancements required in the SS techniques for implementation in the B5G intelligent wireless networks. **Section V** firstly presents the lessons learnt from the detailed survey. Next, the various SS technique issues and challenges existing in the current intelligent wireless networks technology are detailed. Lastly, the open research challenges and potential directions that may be followed are listed. Finally, **Section VI** concludes the article.

## II. Background: SS Techniques

As with every wireless network generation, future wireless networks will also be involved in a quest for new spectrum bands. Notice that there are complex regulations regarding radio spectrum usage, starting from ITU level to the national level spectrum assignment decisions, depending on the country operating the wireless network systems.

### A. Overview of SS

As a consequence of the increased internet users and high bandwidth requirements, there is a spectrum scarcity in the mid-band (1 GHz-6 GHz) that will significantly challenge the 5G/B5G service provisioning [44]. Such a spectrum band is known to be underutilized at different locations around the globe [45]. To minimize this wastage, for each transmission, choice of various bands, following required service/application needs, capabilities/categories of terminal, and conditions of network, can be performed for optimizing bandwidth [46]. Further, owing to this shortage of spectrum in conventional microwave bands (< 6 GHz), there has been a trend of moving the network operations towards a higher frequency band between 30-300 GHz [47]. This is mainly due to the (i) wider bandwidths which can be exploited compared to conventional microwave bands for cellular communications, and (ii) arrays of antennas with high directionality used to provide high gain.

Numerous recent studies have analyzed use of mm-wave for cellular communications [48]. The mm-wave bands can enable traffic transfer to a lesser crowded spectrum from below 6 GHz region. In contrast, low-frequency bands, which have lower capacity, can help maintain the traffic connections even in traffic blockages. The SS techniques are especially relevant when operating in sub-6 GHz bands since, over this range, the frequency slots never increase; however, there is a continuous increase in the traffic demand. Also, recent studies have discussed the need to provision wide band for communication over sub THz frequency which includes the 300 GHz spectrum.

In [49], the authors have presented 300 GHz CMOS wireless transceiver with the capability to transmit data rates upto 80 Gb/s. Further, using statistical models, it has been shown that SE and throughput of network can be significantly improved [50]. Another perspective of SS is in view of radio spectrum management requires (i) a larger framework of the natural resources governance and (ii) multiple governance models at various local and global levels since the available spectrum can be viewed as a resource pool that is common for all the users [51]. The spectrum management model defines multiple rights of property over radio spectrum, and SS introduces such rights, which change depending on sharing model, i.e., SS provides the concrete guidelines to use a shared spectrum pool [52]. The model could be such that it aims to find solutions to the problem of difficulty in excluding, i.e., the complications which prevent others from using the same spectrum. On the other hand, the model may also target spectrum subtractability, which indicates whether the use of spectrum by one user reduces other users' ability to use the same spectrum [53]. Overall, a wise and effective SS between the multiple users of varied radio services will maximize spectrum utilization.

### B. Types of SS Techniques

There could be SS in licensed or unlicensed spectrum bands as illustrated in Figure. 3. For SS in the licensed band, the techniques used may include dynamic sharing, licensed shared access, or spectrum access system [54]. The dynamic spectrum access technique is enabled by advanced scheduling algorithms between the 4G/LTE and 5G, whereas licensed spectrum access uses a centralized database for incumbent protection and a buffer to obtain the spectrum information [55]. Meanwhile, for SS in the unlicensed band, techniques used may include LTE-Unlicensed, LTE-Licensed Assisted Access, MulteFire, and NR-U [56], [57]. The LTE-Unlicensed and the LTE-Licensed Assisted Access are based on the CA implementation to leverage the unlicensed 5 GHz band [58]. The MulteFire techniques do not require an anchor channel which is licensed to utilize the unlicensed band. These techniques enable the promotion of LTE operations in the standalone manner within the unlicensed bands. [59]. With the introduction of 5G NR-U by the 3GPP, 5G NR is now extended to the unlicensed bands via implementation of the Listen Before Talk (LBT) mechanism.

In recent years, studies have revealed the growing business value of the SS techniques implemented in a licensed shared access model. It has also been demonstrated that (i) a significant increase in the capacity gains is achievable via dynamic spectrum access, and (ii) licensed spectrum access enabled SS can be implemented for regulations and standardizations [60]. Further, over the years, multiple standardization efforts have also appeared in regard to the SS techniques, which include the following: (1) standardization for dynamic spectrum access by IEEE Standard Coordinating Committee 41 and IEEE 802.22 [61], (2) collaboration of European Telecommunications Standards Institute and European Conference of Postal and Telecommunication Administration (CEPT) to ensure that licensed shared access standardization output and regulatory framework remain properly aligned and suitably complement one another [62], (3) European Conference of Postal and Telecommunications Administrations (ETSI), which is currently working on an enhanced version of licensed spectrum access to support advanced wireless networks[63], and (4) item Federal Communications Commission (FCC) which has introduced the spectrum access system for enabling shared utilization of Citizens Broadband Radio Service (CBRS) [64].

#### 1) SS Methods

Two key methods exist for SS, viz., Horizontal SS (HSS) and Vertical SS (VSS), which function based on the different spectrum access rights levels given to the primary and secondary users. In HSS, which implements co-primary SS (alternative nomenclature for HSS as per ITU terminology), spectrum is shared between the systems with similar spectrum access rights. In other words, in HSS, all the users have equal rights for utilizing a specific frequency band, i.e., in HSS, multiple CR systems use the same shared spectrum band. Further, HSS refers to the systems which operate with same access rights level, although specific spectrum exploitation priorities may differ. On the other hand, in VSS, comprising primary and secondary users, spectrum is shared between systems with varied spectrum access rights depending on priority.

TABLE III
SPECTRUM ACCESS METHODS IN SS.

| | Vertical | Horizontal | Open |
|---|---|---|---|
| **Description** | Coexistence of primary and secondary CR services | Coexistence of primary CR services with equal (homogeneous) or different (heterogeneous) priorities | unrestricted access |
| **License** | Licensed with overlay, underlay, or interweave-based CR coexistence | Licensed (dedicated to primary users) | Unlicensed |
| **QoS guarantees** | Medium-High | High | Low |

In other words, CR enabled systems perform SS with systems which have no CR capabilities. The CR enabled can utilize spectrum according to regulations imposed by sharing method, i.e., overlay, underlay, or interweave. VSS is also known as hierarchical or primary-secondary SS. There is no mutual exclusiveness between HSS and VSS, as both exist in real-time cases. Such a scenario exists in bands which are unlicensed where users adopting Wi-Fi, Bluetooth, etc., share the spectrum with the same rights, i.e., they implement HSS; whereas, in a few countries, VSS can protect these users in the same configuration via the higher or lower level of spectrum access rights. For instance, in Boston, USA, the feasibility of SS over 100 GHz is demonstrated in [65] via experiments, which implements a real time, dual band backhaul prototype for tracking passive users' presence simultaneously avoiding interference by automatically switching the bands. In general, HSS protects micro licensees from interference within their area of license, while VSS provides protection to incumbents from interference by micro operators. In addition, another method for SS is the Open SS (OSS) technique which offers unrestricted spectrum access to any service/application. Nevertheless, users must still comply with specific rules [66]. Table III summarizes the main distinctive characteristics of the above SS methods.

#### 2) SS Models

Multiple spectrum models can be deployed to implement SS [67], which include: (i) Intra-operator spectrum access model, in which SS is usually the internal decision of the mobile operator within the allotted band; (ii) Spectrum trading or leasing model, in which SS may occur between a new user and the mobile operator as per the rules agreed upon; (iii) Licensed Shared Access, in which VSS protects the primary users and HSS is implemented with potential additional licensed users; (iv) License-assisted access, in which HSS is implemented in unlicensed band as per rules of the band; and (v) Spectrum pooling, in which HSS is used between the mobile operators who participate in the spectrum pooling, and VSS protects the potential users. It must be noted that spectrum models such as, Exclusive Licensed Access also exist, which do not permit any SS between the users. Another spectrum model termed Local licensing (or Micro-licensing) implements HSS between the local licensees as per the defined rules, and VSS is issued to protect them. Note that specialized services based on domain-specific knowledge, such as those required in high demand areas with varied requirements, can be more efficiently provided by local stakeholders than traditional network operators. Therefore, new operator roles are being considered which can (i) offer locally tailored and context specific service delivery and (ii) open mobile market for new entrants [68], [69], [70]. As an example, there is the so-called micro-operator [70], which is basically a stakeholder operating locally, deploys and operates small cell networks, and offers services which are related to context in specified area with/without direct network operator involvement.[1] Currently, opportunities in spectrum for the local 5G networks change significantly among the countries [71] and the stakeholders. Many countries have band slots reserved for local usage via administration [72], while others are not that open. In any case, there is plenty of potential for micro operators to respond quickly to provision locally generated requirements and services requiring high connectivity, especially with assistance of sophisticated network slicing and SS techniques [73], [74].

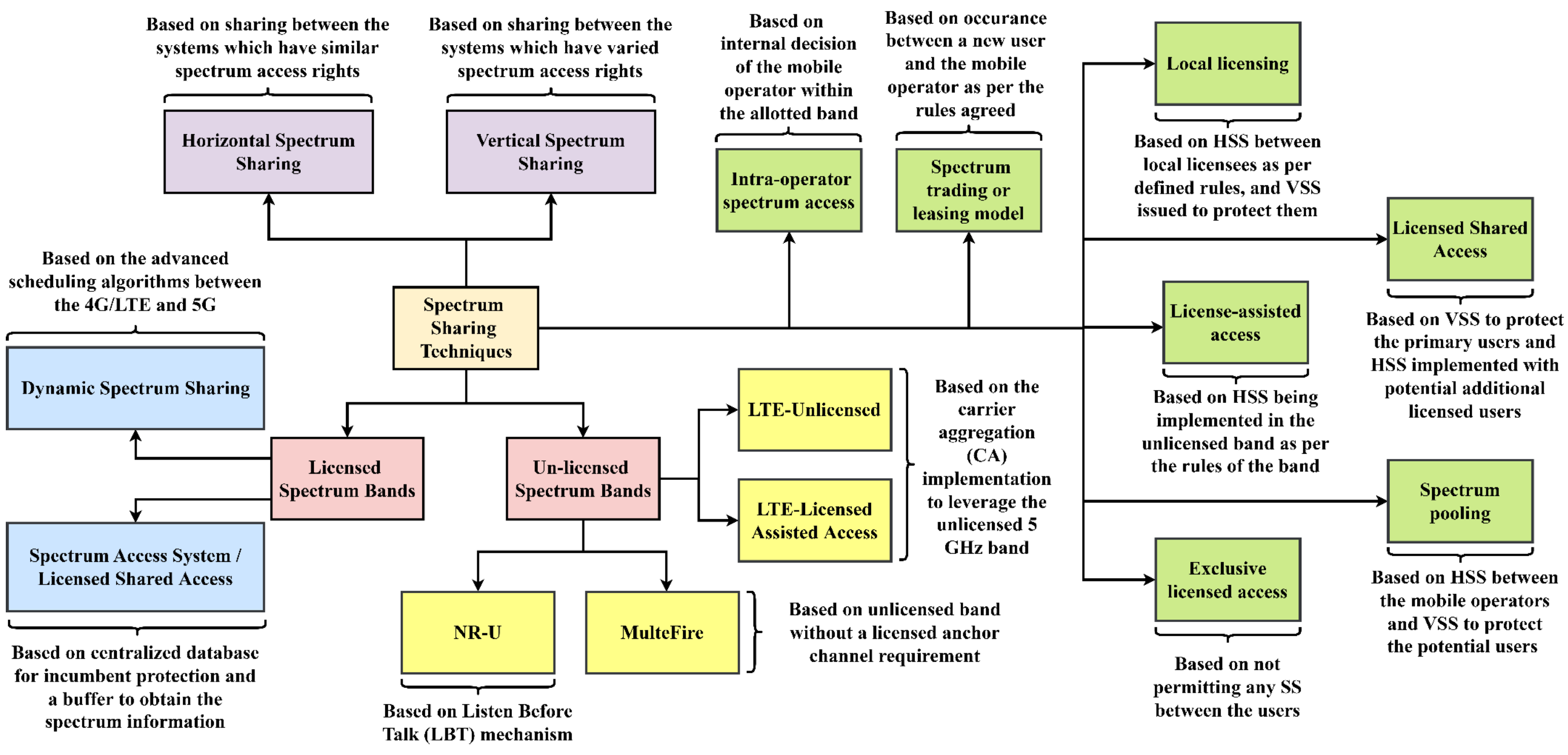

Fig. 3. Types of SS techniques [13], [14].

One solution to spectrum shortage issue is enhancement of use of bands which are available by implementing Dynamic Spectrum Sharing (DSS) techniques in cases when multiple radio systems utilize same band [75]. This implies that one network operator will provide both 4G/LTE and 5G NR services to facilitate DSS between the two services. For example, DSS operates by broadcasting the 4G/LTE and the 5G NR cellular wireless signals over the same frequency, wherein cellular resources between two networks are automatically allotted based on demand. This, in turn, allows mobile network operators to slowly allocate more spectrum resources to the new technology as more users switch to the 5G wireless networks. On the other hand, using DSS technology provides an additional and new frequency spectrum to the LTE network, allowing users to access the network at higher data rates. It must be noted that 4G/LTE and 5G NR operate over different technology frameworks to synchronize across the wireless network. Hence, for DSS to be implemented efficiently, both technologies must cooperate and operate synchronously. When any problem(s) arise, DSS may adopt the multicast-broadcast single frequency network (MBSFN) and non-MBSFN methods to find the appropriate solutions. Also, with advancements in DSS and MIMO techniques simultaneously, the legacy infrastructure can be optimally utilized to save time and cost.

The DSS technology may also allow cellular networks to provide 5G services without physically upgrading their radios. Specifically, the 4G/LTE and 5G networks can use same band and antenna for transmission via the DSS technology. Therefore, significant infrastructure costs may be avoided since it will not be needed to replace the numerous radios over the cell towers in any country as only an update in the software will be required to enable the DSS remotely. This will help usher the ‘5G future’ faster and allow mobile network operators to participate in the 5G technology without purchasing new radios. Lastly, spectrum configuration is also possible in the DSS technology through which details of frequency, band plan, block size, address, and specifications can be obtained, which aids in the identification of the characteristics of that spectrum block being auctioned from an imaginary spectrum pool.

[1]Note that the local 5G network is deployable by traditional network operators which permits novel business opportunities to them.

### C. Challenges in SS

The spectrum utilization of the available lowest bands depends on both frequency band and the data rate requirements. Hence, the more efficient the frequency sharing is, the more optimized the spectrum utilization becomes, which leads to a better overall user experience. However, much research is still needed in this regard owing to the current spectrum usage trends in various bands. The numerous related challenges are as follows:

1) Notice that the research community has extensively proposed, studied, and analyzed CR techniques for SS, enabling multiple radio systems operating in same band. However, developing SS techniques exploiting the CR technology requires extensive discussions and further research for implementation.
2) The complex issues of reassigning specific frequencies and purchasing the additional frequency spectrum require much research.
3) Multiple SS methods have been proposed which are centralized in nature; however, these techniques are not scalable and require very high computational resources. Hence, rigorous research on decentralized SS techniques is needed to be conducted.
4) Issues such as high administrative cost, unoptimized spectrum allocation, possible illegal uses of unused spectrum via CR techniques, multiple security attacks, and co-existence issues require further investigations.
5) It is required to investigate the capital expenditure-related issues since operators will need to update their equipment. The operational expenditure will also need to be analyzed for small cell deployments. For network operators, assessment of the return on investments will be a major issue owing to increase in network users over a period of time. There is also a fundamental need for appropriate pricing models since primary users and network operators demand decent economic profits. Hence, it is imperative to introduce incentive base models to ensure that both parties participate in SS mechanism.
6) To implement the dynamic spectrum methods, the network's critical components will need to be synchronized and coordinated. Hence, the effect on performance of network owing to signaling data from such coordination needs to be investigated.
7) With increase in demands from network operators to cooperate in the SS process, it can be anticipated that the SS framework will become more complex, thus, motivating further extensive research.
8) Given the SS contract, primary users may have a much smaller spectrum footprint geographically, lesser flexibility regarding spectrum utilization, and increased confinement in their utilized times. Also, the services of the primary users may be degraded if the network operators are not adhering to the performance levels that were negotiated a-priori. In turn, this may limit the network operators from applying the less reliable shared spectrum to meet the diverse service requirements. This aspect needs to be investigated.
9) In cases which relate to the implementation of NR-U, the hidden and exposed node issues arise owing to to beam enabled transmissions, which require further investigation.
10) Numerous standardization concerns remain, such as the time necessary to evacuate the licensed spectrum access licensees from the spectrum band and the appropriate steps to execute when other incumbents do not wish to share the spectrum appearing in the given spectrum band.

## III. SS Techniques for existing Intelligent Wireless Networks

In the previous wireless network generations, spectrum usage was primarily exclusive. Specifically, network operators acquired spectrum licenses from national regulators through spectrum auctions. As the network operators had already cleared the spectrum bands for primary users, SS was not required. Specifically, operators paid to obtain the license, and users implemented different techniques to obtain the spectrum access rights, which did not require SS between the wireless systems.

Simultaneously, with advancements in the wireless networks, more spectrum bands have opened for operation. E.g., in 2019, for the International Mobile Telecommunications (IMT) systems, 24.25–27.5GHz, 37–43.5GHz, and 66–71GHz mm-wave bands are made available. However, the key challenge is efficiently utilizing such spectrum bands, which offer enormously large bandwidths; hence, SS attains importance. The SS mechanisms can be implemented through [76]:

1) permitting a network operator for 5G migration using spectrum band wherein it can implement the existing 3G/4G systems. This will facilitate a flexible interworking between the various technology generations within a network operator's system, or
2) assigning local 5G spectrum licenses, which involves an inter-operator SS model to perform spectrum re-farming as a solution to divide the spectrum between the various wireless network generations, or
3) developing 5G system variants for unlicensed spectrum access. This will resolve the key challenge of adopting such mechanisms, which enable various technologies in coexisting with other technologies in a fair manner within the unlicensed spectrum.

It must be noted, though, that SS levels will vary depending on the implemented model. In the future, the B5G wireless networks will encounter extremely dynamic environments during operations. In such cases, finding a new licensed spectrum will be very challenging due to the immense rivalry between the users for the spectrum bands (e.g., used for satellite and broadcasting); hence, SS will become a necessity. The SS techniques will also play an essential role in the B5G systems as they will address the challenges concerning the protection requirements of the existing users.

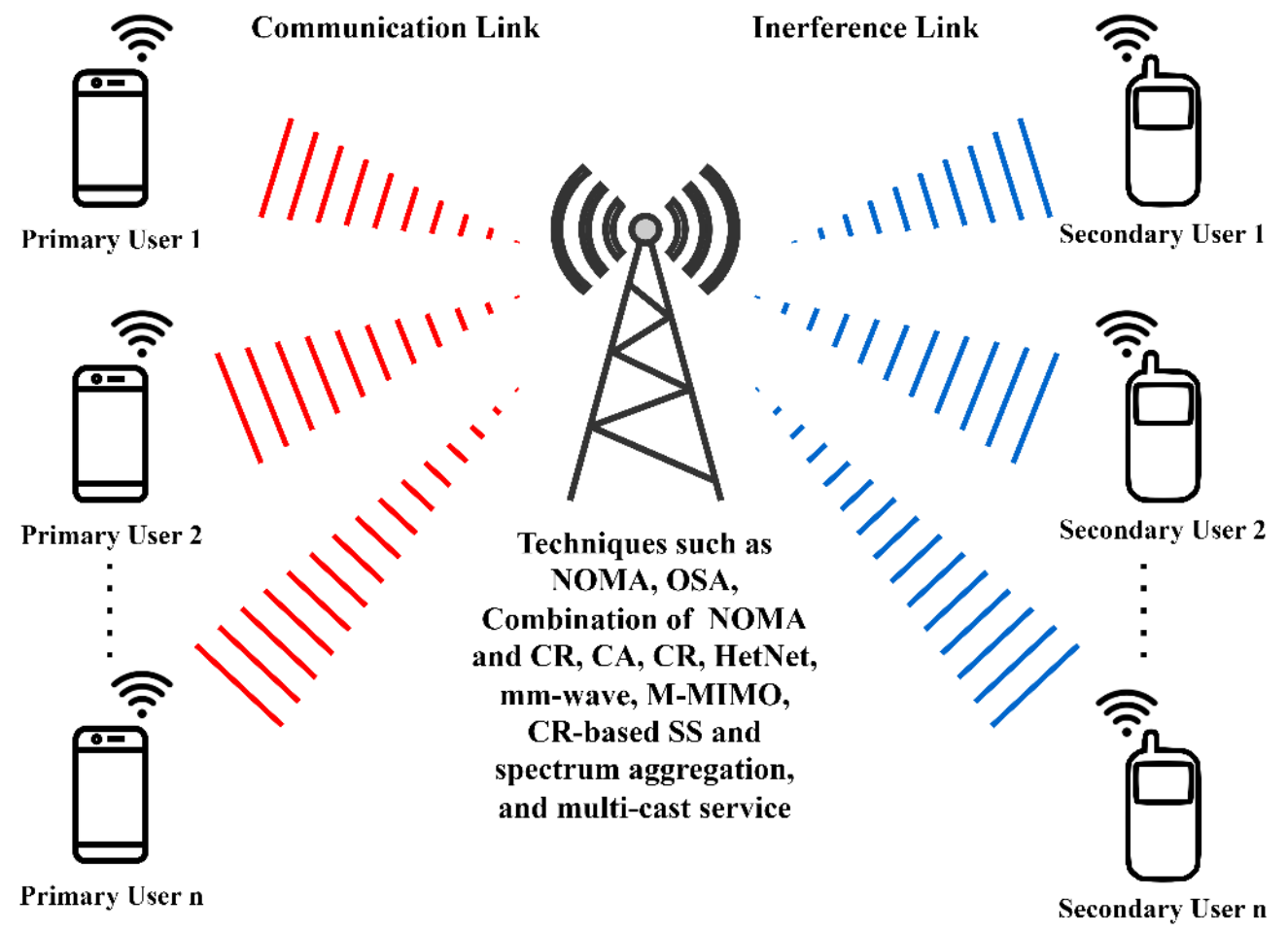

Fig. 4. SS in a CR environment.

Previous discussions make it evident that the next-generation intelligent wireless network will require advanced and efficient SS techniques. In recent years, the research community has identified multiple technologies that can be used to advance the SS techniques for intelligent 5G wireless networks. In the sub-sections which follow, we segregate related studies according to technology used for implementing the SS technique.

### A. SS using CR

SS is effective for enhancing spectrum use in a CR environment. As shown in Figure. 4, cognitive sharing of the licensed users' spectrum enables the unlicensed users to exploit such spectrum resources, which are rigid, simultaneously ensuring that the performance of the licensed users is not degraded.

#### 1) Related Works

In [2], the authors surveyed the most recent SS technologies for facilitating the deployment of 5G networks. The SS techniques survey is categorised depending on architecture of network, behaviour of spectrum allocation, and method of spectrum access. Further, study of CR technology in SS for 5G implementation is performed. Thereafter, discussions are presented on issues and challenges in existing implementation of SS and CR. Lastly, methods for supporting 5G advancement are detailed. Also, survey on CR technology for SS in 5G networks is conducted, followed by detailed discussions on issues and challenges. Finally, the authors have discussed the technology for advancing the 5G wireless networks.

In [14], the authors described the fundamental idea of non-Orthogonal Multiple Access (NOMA) and the relation to CR networks. The authors state that by including the CR technology with NOMA, there will be an evident improvement in SE and EE compared to implementing the existing Orthogonal Multiple Access (OMA) technologies. With implementation of NOMA, information intended for the primary user is decoded by secondary users, and thus secondary users will operate as relays. This will provide an opportunity to access the primary user's channel for performance improvement of CR-NOMA system. Further, the authors presented a cooperative method for transmitting to achieve enhanced performance in presence of higher number of secondary users participating to relay simultaneously, providing maximum diversity. As a limitation, the proposed approach uses omnidirectional antennas, and secondary users cannot identify primary users' exact location.

The authors in [21] surveyed the state of art techniques proposed to cater to multi-cast services in the CR wireless networks. The survey details different SS models adopted in the CR wireless networks to fulfill the main aim of multi-cast services, viz., end-to-end communication. The authors have presented the multi-cast traffic nature, the key techniques and methodologies which can be implemented to enhance the efficiency of multi-cast services, and techniques to minimize the interference experienced by the incumbents. The authors have also provided possible design guidelines which can be adopted for multi-cast services in the 5G networks. Finally, key research issues have been identified, and relevant future research directions have been discussed.

The authors in [77] proposed various SS approaches using the CR network's Opportunistic Spectrum Access (OSA) models. The methods adopted are Markov Decision Process (MDP), Multi-armed bandit, optimal stopping priority, and game theory. Operations to accommodate more users and increase the SE, such as parallel sensing, are implemented by the optimal stopping priority method. Various MDPs, such as partially observable MDPs, constrained MDPs, and discrete-time MDPs, have been studied and extensively used to analyze the sequential decision-making in SS. In addition to the MDPs, game models such as static, repeated, graphical, evolutionary, and coalition models have been used for throughput maximization and interference avoidance. Lastly, it is shown that a fixed set of models can be sensed through the game theory approach.

The authors in [78] analyzed throughput of CR network by considering factors such as random sensing and energy queue for multiple secondary users. It is shown that the sensing probability of the spectrum decides the throughput. The authors highlighted that in the proposed spectrum sensing approach, saturated sources and traffic of primary users are not considered. Further, considered network is a multi-user CR which accounts for energy harvesting. Also, in this model, multiple secondary users share a common primary channel according to a two-step opportunistic spectrum access mechanism, which comprises sensing of random channel followed by accessing of random channel. This two-step opportunistic spectrum access model has two adjustable parameters viz., probability of channel sensing $q$, and probability of channel access$p$. The authors have derived the network throughput considering capacity of energy queue tending to infinity or zero, and have investigated access probability impact on highest throughput which can be achieved for case of an infinite capacity. The result indicates that the access probability has a minor effect on maximum achievable throughput when channel sensing probability is chosen optimally. It is also demonstrated that the maximum achievable throughput is the same for the two extreme cases if the sensing and access probability are chosen optimally. Lastly, the authors state that even though this result is obtained by investigating two extreme cases, it is expected to hold for any energy queue capacity.

In [79], the authors highlighted that NOMA and CR together could solve the problem of shortage of particular frequency in addition to provisioning additional benefits in terms of spectrum. Further, NOMA and CR together can achieve the 5G requirements of high performance, augmented connectivity, and least delay. The study recommends that NOMA and CR together can effectively use the wireless spectrum with determination of sharing method. The authors stated that integrating NOMA with CR needs additional analysis for the related implementation.

The authors in [80] considered a CR network with multiple secondary users equipped with energy harvesting capabilities. The authors investigated joint impact of probability of sensing, capacity of energy queue, and access probability on highest throughput which is achievable in a multi-user CR network. It is shown that channel access probability does not affect highest throughput which can be achieved if probability of sensing the channel is selected appropriately. The authors derived the probability of optimal sensing in terms of network parameters such as channel availability probability and contending secondary users amount. Further, results are expected to hold for any energy queue capacity. However, the study does not confirm the results for unsaturated resources.

The authors in [81] reviewed cooperative communication and techniques such as CA, CR, and small cell enabled Heterogeneous Networks (HetNet), high-spectrum access, and M-MIMO, for attaining the results which are desired. The authors have highlighted potential issues associated with management of spectrum and have detailed the possible solutions. Advanced user applications, less delay, and large throughput for increased communication efficiency is envisioned to be supported by B5G networks communication. The authors state that there is a need to formulate advanced SS methods to attain complete spectrum capacity for existing 5G networks. Hence, managing spectrum will be key to achieve this aim; on the other hand, multiple problems arise in the design of such future networks. The five key topics addressed in the article are high-spectrum access, CA, small cell networks, M-MIMO, and CR. The authors conclude that using various methods simultaneously in existing 5G network is necessary for enhancing overall SE.

Considering wireless networks enabled by CR technology, the authors in [82] focused on issue of resource allocation via the sensing-based interference recording method. In the study, the base station is considered to control the users' transmission through such channel allocation mechanisms, which are dependent on recorded interference data. The results demonstrate that proposed model effectively records statistical behaviours of CR network interference and allocates the channels to attain higher capacity of system.

In [83], the authors proposed novel CR-based SS and spectrum aggregation techniques for 5G networks. The proposed methods exploit licensed spectrum with primary user networks and unlicensed spectrum collected via ISM bands. The authors also included the dynamic spectrum management process within the proposed method and ensured that harmful interference to the incumbents is controlled. Also, the critical trade-off between efficiency of sharing and efficiency of aggregation is addressed. Depending on this trade-off, a spectrum lean-management technique and water-filling algorithm are proposed to aid in dynamically accessing the spectrum. Extensive simulation results demonstrate that proposed technique improves system performance.

*2) Summary*

To summarize, the CR technology provides an efficient method to optimize the spectrum. Specifically, the use of CR technology in conjunction with NOMA improves the SE and EE as secondary users can decode information intended for the primary user to solve problem of scarcity of a specific frequency spectrum. Further, to improve the system throughput, a cooperative transmission scheme including several secondary users can be implemented, together with a dynamic SS approach enabling the sensing of more channels.

The sensing probability of the spectrum is paramount in deciding the system throughput as it enables efficient sensing of the spectrum. However, it has been shown that the channel access probability may not affect the maximum achievable throughput if channel sensing probability has been selected appropriately. The implementation of sensing-based interference recording for wireless networks enabled by CR for resource allocation has also been found to be effective. Such a scheme is appropriate for recording CR wireless network interference's statistical behavior and allocating channels to attain a higher system capacity. The CR-based SS and spectrum aggregation techniques, including dynamic spectrum management, have also been shown to improve system performance.

Overall, CR is appropriate for intelligent wireless networks, wherein there are fluctuations in the available spectrum and policies and heterogeneous QoS demands according the applications. Table IV summarizes all the recent studies that have implemented SS using the CR technology.

### B. ML oriented SS

The rapid development of advanced wireless technologies has resulted in the requirement for increased spectrum resources. The large available spectrum must be effectively utilized in an intelligent manner. Hence, it is now mandatory to introduce learning and reasoning capabilities to implement SS within the 5G/B5G networks (see Figure. 5). Implementing the ML techniques adds new a dimension of intelligence within the existing wireless networks.

*1) Related Works*

The authors in [84] detailed the operation of of the macrocell and the femtocell structure. The relation between mean capacity of each user and number of users is demonstrated. Further, cooperative bandwidth sharing is proposed using the game theory method, showing that resource utilization is optimized using the cooperative game between different cells. It is also demonstrated that cooperative bandwidth sharing achieves better performance with more subscribers. However, the study

TABLE IV
SUMMARY OF SS TECHNIQUES USING CR TECHNOLOGY.

| Reference | Aim | Methodology | Key Results | Limitations |
|---|---|---|---|---|
| [77] | • Presented strategies and solutions for distributed channel selection in regard to opportunistic spectrum access. • Detailed various SS approaches in opportunistic OSA models. | • Game theory model, Markov decision, optimal stopping, and multi-armed bandit based spectrum allocation approaches are used. | • The evaluated metrics include throughput and EE. • Among the four proposed models, it is found that each method is superior with respect to one parameter in OSA-based sharing. | • Game theory approach provides sensing of only a fixed set of models, whereas dynamic scenario contains many channels. |
| [78] | • Investigated joint impact of sensing and access probability, and energy queue capacity on maximum achievable throughput in a multi-user CR network incorporating energy harvesting. • Analyzed throughput of the CR network. | • Analysis of throughput of CR network by considering factors such as, random sensing, energy queue for multi-secondary users, and accessing is presented. | • The metrics evaluated include energy arrival rate, channel availability probability, and the number of contending secondary users. • Designed Energy Harvesting model for multi-user CR network. | • Saturated sources and traffic of primary users are not considered. |
| [14] | • The fundamental idea of NOMA is detailed, which boosts SS. • Presented a cooperative transmission scheme for efficient SS. | • Existing queuing-based spectrum approach, Spectrum hand-off, and Dynamic Spectrum access techniques are compared. | • Network throughput and power are considered as the performance metrics. • The proposed method achieves better performance when more secondary users participate in relaying with maximum diversity. | • Proposed approach uses omnidirectional antennas; however, the secondary users cannot identify exact location of primary users. |
| [79] | • Compared NOMA and CR techniques to share spectrum efficiently. • Demonstrated that combination of NOMA and CR can solve the problem of scarcity of a particular frequency spectrum and provide additional frequency benefits. | • NOMA integration with CR is proposed. | • Combining NOMA and CR can also meet 5G standards for high performance, excellent connectivity, and low latency. • The study recommended that wireless spectrum be used more effectively by combining NOMA and perceptual radio techniques with the determination of sharing method. | • Integration of NOMA with CR needs more analysis for implementation. |
| [80] | • Investigated joint impact of sensing probability, energy queue capacity, and access probability on maximum achievable throughput in a multiuser CR network. • Prediction of channel availability in multiuser CR and network using sensing probability and energy rate calculation are presented. | • Derivation of optimal sensing probability as a function of network parameters such as, channel availability probability, energy arrival rate, and a number of contending secondary users is detailed. | • Blocking probability, service rate, and arrival rate are considered as the performance metrics. • It is shown that channel access probability does not affect maximum achievable throughput if the channel sensing probability is chosen appropriately. | • Results are expected to hold for any energy queue capacity. • The study does not confirm results for unsaturated resources. |
| [81] | • Provided an in-depth review of potential problems associated with spectrum management and available cooperative communication techniques proposed. • Discussed CR to attain a high level of spectrum resource optimization required for current 5G networks. | • AI/ML based channel characteristics prediction methods are detailed. | • Study has shown that AI and M-MIMO-based methods can achieve adequate QoS performance for high altitude users. | • The reviewed AI/ML concepts are still not realized. |
| [2] | • Presented detailed survey on most recent SS technologies which enable 5G networks. • The SS methods are classified, following which a detailed review is conducted on the SS techniques' related studies | • Placement of studies included in survey within the categories is based on behaviour of allocating spectrum and methods used to access spectrum. • A detailed survey of SS CR technology in 5G networks is conducted. • Detailed discussions on issues and challenges in the SS and CR implementation for 5G network is presented. | • The considered performance metrics are usage of spectrum, consumption of energy, and cost efficiency. • The technology is identified which will be implemented to advance 5G wireless networks. | • Discussions are conducted on issues and challenges in regard to implementing only the SS and CR technologies. • Other aspects are not covered. |
| [82] | • Effectiveness of a wireless network enabled by CR is analyzed. • The issue of resource allocation enabled via the sensing-based interference recording for wireless networks enabled by CR is considered. | • The base station is considered to control users' transmission through channel allocation mechanisms based on the recorded interference information. • Such information is recorded, and a low-complexity allocation method is introduced to increase task capacity. | • The study includes a number of valid channels as the performance metric. • The proposed model effectively records statistical behavior of CR wireless network interference and allocates the channels in view of higher system capacity. | • The proposed technique is not applicable to advanced wireless networks. • A new method requires to be implemented. |
| [83] | • Performance of a CR-enabled dynamic SS technique is analyzed. • Novel CR-based SS and spectrum aggregation techniques for 5G wireless networks is proposed. • Dynamic spectrum management process is included within the proposed method. | • The proposed methods exploit licensed spectrum with primary users and unlicensed spectrum through secondary users. • Interference to incumbents is controlled. • Key trade-off between sharing- and aggregation-efficiency is addressed based on which spectrum lean-management scheme and water-filling algorithm are proposed, which aid in dynamically accessing spectrum. | • Sharing and aggregation efficiency are considered as performance metrics. • Simulation results show that proposed technique can provide system improvement. | • The proposed technique is implemented for a 5G network scenario. • The applicability to B5G networks is not considered. |
| [21] | • Presented a detailed survey on state of art techniques proposed to cater to multicast services in the CR wireless networks. • The survey details different SS models adopted in CR wireless networks, which are implemented to fulfill the main aim of multicast services, viz., end-to-end communication. | • Presented multicast traffic nature, key techniques and methodologies which can be implemented to enhance efficiency of the multicast services, and techniques through which interference experienced by incumbents can be minimized. • Provided possible design guidelines which can be adopted in view of multicast services in the 5G networks. | • Identified critical research issues and discuss relevant future research directions. | • The survey is restricted to 5G wireless networks. |

is limited in regard to (i) cell-to-cell interference being high, (ii) the proposed approach being more feasible for indoor communication in view of distance, and (iii) users trying to connect with two cells, which creates more complications in regard to bandwidth sharing.

In [85], a Transform Domain Pre-coding (TDP) method is proposed to design the pre-coder and feedback to solve the issue of spectrum sparsity. Link-level simulations analyze and evaluate realistic parameters such as windowing in frequency domain, hybrid beamforming, and allocation of power.

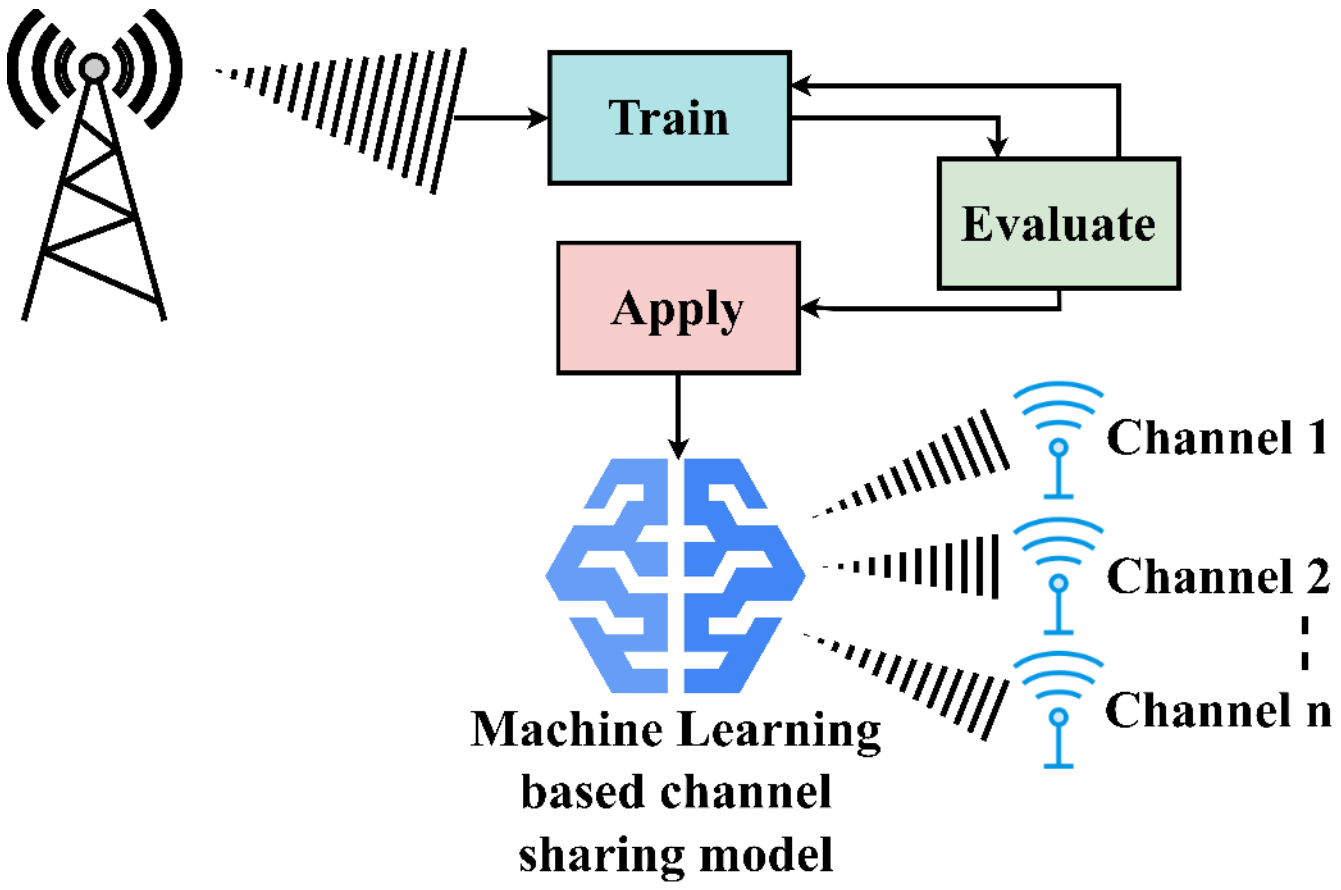

Fig. 5. SS using ML techniques.

Results indicated that 2.38% $\sim$ 21% SE enhancement and 60% $\sim$ 89% overhead reduction can be achieved by TDP. In practical systems, TDP is associated with frequency domain windowing and hybrid beamforming, which are selected for exploiting sparsity of channel in the transform domains. Firstly, as opposed to similar latency for different antennas in complete digital pre-coding, multiple latency is chosen for different beams within hybrid beamforming. Secondly, quantized latency is assumed for orthogonal frequency division multiplexing (OFDM) systems. Next, windowing in frequency domain due to lesser assigned sub-carriers amount than Fast Fourier transform (FFT) size affects are observed for channel sparsity in delay domain. To cope with issues of windowing in frequency domain and optimization problems on quantized letency, compressive sensing (CS) method is implemented for designing TDP delays. In addition, TDP considering unequal and equal power allocation among sub-carriers is formulated. The proposed TDP is shown to perform better compared to current feedback overhead and SE methods.

In [86], the authors discussed the MDP and the game models and demonstrated that the most eye-catching asset of the MDP models is that Markovian dynamics of the spectrum scenarios can be efficiently modeled and analyzed. Also, authors have pointed out that measured spectrum scenario in opportunistic spectrum access systems is non-reactive, i.e., actions of secondary user will not affect system's evolution. It is also validated that events of primary users is approximately characterized by discrete-time or continuous Markovian processes. Hence, the MDP models can be considered efficient and promising OSA system solutions. It is then shown that the current MDP models for varied OSA cases are better considering sensing ability, sensing strategy, action, optimization objective, etc. In view of the game model, the authors have demonstrated that capturing the interactions among multiple secondary users will be efficiently conducted by the game models as they provide a learning algorithm to predict the system's steady states. It also allows the analytical characterization of the performance of the steady states. The authors state that a flexible design is possible by game models. Therein, a utility function determines structure and steady state of game; hence, it is a key factor in the game models. The game models are useful in solving issues such as adapting the rate, controlling power in distributed manner, and controlling access for multiple users. As for limitations of the study, the authors state that (i) almost all game models use sensing methods which are parallel, i.e., they sense fixed channels set within a slot, and such strategies, although convenient for analysis, lead to conservative throughput, (ii) game models rely on knowing channel's statistical information a-priori; however, availability of such information at all times is not possible which makes current game models not implementable when statistics information is unavailable, and (iii) MDP models show suitability for individual secondary user systems instead of multiple secondary user systems since for obtaining optimal policies of MDP model; however, it needs stationary scenario.

In [87], the authors detailed the aspect of channel sensing by secondary user. The sensing mechanism of the two channels in an OSA system is discussed, which is then evaluated considering the throughput. Markov-chain-based greedy channel assignment shows that secondary user can sense two channels and use unutilized spectrum. The results demonstrate that the secondary user attains maximum throughput of 0.45 Mbps and can exploit more resources for data transmission. Further, the secondary user is also shown to incur a minimum of 1.2 seconds, considering 10 users. Lastly, the authors showed that the secondary user can exploit more spectrum resources.

The authors in [88] proposed an innovative method that permits the Wi-Fi networks and LTE-unlicensed networks to exist together within same unlicensed spectrum of 5G band. The issue of augmenting Quality of Experience (QoE) of LTE–U is solved using game theory. Specifically, the authors solved unlicensed band choice issue of small cell base stations through coalition game. The co-existence issue is solved in each coalition through the Kalai-Smorodinsky bargaining solution (KSBS), whereas RA problem of every small cell base station is solved using learning game. The results demonstrate that proposed method generates good fairness score in comparison to other method and is better to manage Wi-Fi systems than LBT. As a limitation, the proposed approach performs efficiently only in indoor communication applications.

The authors in [89] investigated the trade-off between two performance metrics considering static and mobile communication. Depending on expressions, complexity of computation used for configuring hybrid precoder is minimized. This is utilized for adaptively activating the needed RF chains for specific MIMO system and state of channel. Results show that a specific amount of RF chains must be activated for maximizing EE at high signal-to-noise ratios (SNRs), which is usually different from optimal configuration to increase SE. In addition, for lower SNRs, it is demonstrated that simple analog beamforming, which utilizes one RF chain, is optimal for SE and EE. Also, proposed mobility aware hybrid precoding effectively achieves beamforming gain between high speed mobile devices.

In [90], the authors presented a detailed survey on ML-enabled SS techniques, critical issues concerning security, and related mechanisms of defense. Specifically, the authors elaborated on the state of art techniques which can be adopted to enhance SS-based systems' performance and presented the security-related challenges arising from physical layer and the related solutions enabled by ML techniques. Lastly, the authors also presented detailed discussions in regard to the research issues on ML enabled SS methods.

The authors in [91] proposed a novel SS technique for supporting massive number of radio networks. Moreover, the authors proposed a 5G amenable novel service-enabled network management architecture for integrating ML techniques for radio resource management and SS. Lastly, the authors also proposed a new 3-stage ML framework for exploiting forecasts, clusters, and reinforcement learning (RL) methods to implement proposed SS technique. It is shown that proposed solution can permit large amount of arbitrary 5G network slices for effectively and dynamically sharing the spectrum amongst one another or with alternate networks.

In [92], the authors overviewed various SS levels and methods in the literature. Additionally, the authors discussed the potential offered by the different SS methods. Lastly, the significant role of ML techniques in facilitating automation and dynamic SS is also detailed.

The authors in [93] presented a dynamic access aware bandwidth allocation method that meet every cell's dynamic traffic requirements, following which, it allocates the required bandwidth from a shared pool of spectrum. Further, in view of acquiring knowledge regarding the needs of the access network, the authors presented an ML-enabled method for predicting the network state and managing the available spectrum efficiently. The proposed approach is compared with two existing techniques considering spectrum allocation accuracy and utilization efficiency as the performance metrics. The results show that ML-enabled technique can outperform current techniques and achieve performances close to an ideal dynamic system.

#### 2) Summary

To summarize, the ML techniques can be implemented to address the complex interactions among the co-existing networks. These methods will be able to tackle the high level of uncertainty via learning from the observations and can also appropriately model complex interactions. In specific, resource utilization can be optimized by implementing a cooperative game between the various cells. In this regard, it has been shown that a cooperative bandwidth-sharing method achieves better performance with more subscribers. Further, TDP-enabled SS methods have been observed to perform better compared to existing techniques in terms of SE. Also, the SS mechanism can be improved by implementing the MDP models, which have demonstrated enhanced sensing ability, efficient sensing strategy, related actions, optimization, and objective. Game models can capture the interactions among many secondary users, efficiently providing learning algorithms for predicting the system's steady states. These models have been shown to allow analytical characterization of the steady state(s) performance and help solve issues such as adaptation of rate, distributed power control, and multiple access control. Also, mobility aware hybrid precoding improves SS and can effectively achieve beam-forming gain between high speed mobile devices. The RL-enabled SS methods permit a large number of arbitrary network slices, which enables sharing of the spectrum effectively and dynamically among the same radio networks or with other radio networks. Such methods have been shown to outperform existing techniques and achieve performances close to an ideal dynamic system.

Overall, the ML techniques will leverage learning within the wireless networks, which will aid in addressing complex interactions between the co-existing networks. Table V summarizes all the recent related studies which have implemented SS using ML technology.

### C. Blockchain oriented SS

Distributed Ledger Technologies (DLTs) are implemented as a network of nodes connected in Peer-to-Peer (P2P) manner and advocate distributed storage of digital ledger. This implies that participating nodes in the P2P network maintain an exact replica of the digital ledger which is updated in a synchronized manner [94]. Moreover, DLT uses consensus mechanism and cryptographic techniques (such as asymmetric keys, hashing, digital signature, and Merkle tree) to make the system secure and reliable even in the presence of faults and malicious nodes.

Blockchain technology is the most prominent type of DLT and has received all-around attention from industry and academia [95]. Blockchain follows a typical data structure for creating and updating distributed ledger [96]. As the name suggests, blockchain comprises a tamper-proof sequence of blocks where each block contains a set of valid transactions. To maintain chronological order of the blocks, these are chained (or connected) together using a cryptographic hash-based chain. Every new block is chained with the most recent block in ledger by establishing a cryptographic hash link between the two blocks.

The smart contract is a computer program that sofwarizes all terms and conditions of a negotiated agreement between non-trustworthy multiple parties [97]. Once deployed on the top of a blockchain, it becomes tamper-proof and transparent to the participating members, establishing trust in the system. Furthermore, smart contract can be designed with predefined conditions that enable the automatic execution of that intelligent contract when met in the future. Thus, a smart contract helps in automation, trust-building, and enforcement of negotiated agreements.

Blockchain and smart contracts are suitable for many domains and sectors, such as healthcare, supply chain, education, and public governance. In the same direction, blockchain is envisioned to play a pivotal role for mobile networks [98], [99]. Among others, decentralized and secure management of resources is one crucial area where blockchains are to be leveraged [100]. This section, in particular, discusses the use of blockchains for SS.

#### 1) Related Works

Authors in [101] advocated using blockchain and AI for intelligent management of 5G networks. The authors considered an open business model that allows (i) dynamic spectrum leasing by mobile network operators (MNOs) from a governmental organization that owns the spectrum and (ii) easy entry for a new entrant to become MNO by leasing resources

TABLE V
SUMMARY OF SS TECHNIQUES USING ML TECHNOLOGY.

| Reference | Aim | Methodology | Key Results | Limitations |
|---|---|---|---|---|
| [87] | • Spectrum access with two-channel sensing in CR networks is presented.<br>• The sensing mechanism of two channels OSA system is discussed and evaluated | • Markov-chain-based greedy channel assignment is used. | • Sensing time and throughput are considered as performance metrics.<br>• The secondary user achieves a maximum throughput of 0.45Mbps and can exploit more resources for data transmission. | • The aspect of secondary users exploiting more resources is not addressed. |
| [88] | • Maximizing QoE of LTE–U system and solving unlicensed band selection problem of the SBSs is addressed.<br>• Proposed an innovative method that permits Wi-Fi networks and LTE-U to coexist in same unlicensed spectrum of 5G band | • QoE issue is solved using the game theoretic approach, and the co-existence issue is resolved in each coalition using KSBS.<br>• Q-learning solves resource allocation problem of each SBS. | • The considered metrics include system efficiency and fairness.<br>• The proposed method provides good fairness score in comparison to other methods.<br>• The proposed approach is better for managing Wi-Fi systems compared to basic LBT. | • Proposed method performs only in indoor communication applications. |
| [84] | • Analyzed the SE and compared two game models.<br>• Built two game models to optimize RA corresponding to different situations. | • Bandwidth sharing game i proposed for efficient bandwidth sharing by cooperation between different cells such as, femtocell and macro cell. | • It is shown that as number of subscribers increases, cooperative allocation attains better performance.<br>• The models are evaluated considering SE as the performance metric.<br>• SE is greater in three-layer heterogeneous network compared to two-layer networks. | • Cell-to-cell interference is high.<br>• The proposed approach is more feasible for indoor communication in view of distance.<br>• If users try to connect with two cells, then more complications occur in bandwidth sharing. |
| [86] | • Identified essential components, fundamental trade-offs, and practical constraints in opportunistic spectrum access.<br>• Systematically tackled optimal integrated design and quantitatively characterized interaction between signal processing for opportunity identification and networking for exploitation. | • Decision-theoretic framework based on partially observable Markov decision processes theory is presented. | • The considered performance metrics are probability of detection and false alarm.<br>• The framework allows for systematic tackling of optimal integrated design.<br>• The framework also permits quantitatively characterizing interaction between signal processing for opportunity identification and networking for opportunity exploitation. | • It is shown that parallel sensing strategies are more prominently used in game models, i.e. in given slot, fixed set of channels are sensed.<br>• These strategies are aimed to ease analysis; however, at the cost of conservative throughput.<br>• Game models rely on knowing statistical information of channel a-priori; however, understanding this at all times is impossible, making traditional game models not workable without statistics information.<br>• For Single secondary user systems, MDP models are more suitable than multiple Secondary user systems due to the fact that MDP model needs environment to be stationary to obtain optimal policies. |
| [85] | • Observed scarcity of channels and allocation.<br>• Proposed design of precoder and feedback | • TDP is implemented. | • To evaluate performance, hybrid beamforming, frequency domain windowing and power allocation are used as the metrics.<br>• Link level simulations show 60% - 89% overhead reduction and 2.38% - 21% SE enhancement achievable by TDP. | • Improvement in precoder for B5G is still an open challenge. |
| [89] | • Investigated trade-off between two performance metrics and selected number of active RF chains via closed form expressions in scenarios involving static and mobile communication scenarios.<br>• Configurations of a flexible hybrid precoding scheme for mm-wave MIMO communication, where a number of active RF chains can be adjusted to achieve optimal, yet different, settings for two performance metrics is detailed. | • Mobility-aware hybrid precoding is shown to achieve beamforming gain between high-speed mobile devices effectively. | • Results are obtained in terms of SE, EE, and SNR.<br>• It is observed that to maximize EE at high SNRs certain number of RF chains need to be activated which is generally different from optimal configuration to maximize SE.<br>• It is observed that for low SNRs, simple analog beamforming, with single RF chain, is optimal for both SE and EE. | • Configurations have reasonable power consumption for high-speed mobile communications. This needs investigation. |
| [90] | • Presented a detailed survey on ML-enabled SS techniques, critical issues concerning security, and related mechanisms of defense.<br>• Elaborated on state of art techniques which can be adopted to enhance performance of the SS-based systems. | • Presented security-related challenges arising from physical layer and the solutions enabled by ML techniques. | • Presented detailed discussions in regard to research issues on ML-enabled SS techniques. | • It is shown that SS technique is efficient in enhancing spectrum utilization.<br>• The survey is restricted to 5G wireless networks. |
| [91] | • Introduced RL technique for SS in 5G network.<br>• Proposed a novel SS technique for supporting a very large number of radio networks.<br>• Proposed a 5G amenable novel service-enabled network management architecture for integrating ML techniques for radio resource management and SS. | • Implemented a new 3-stage ML framework for exploiting forecasts, clusters, and RL methods to implement the proposed SS technique. | • SE is considered as the performance metric.<br>• The proposed solution can permit a large amount of arbitrary 5G network slices in accordance of effective sharing of spectrum dynamically with one another or with others radio networks. | • The study is limited to 5G networks scenario and needs advancements to be implemented in the B5G networks. |
| [92] | • Provided an overview of the various SS levels and methods in the literature.<br>• Discussed the potentials offered by the different SS methods. | • Detailed the significant role of ML techniques in facilitating automation and dynamic SS. | • It is shown that dynamic SS in the 5G networks results in efficient spectrum utilization. | • The study only focuses on 5G networks. |
| [93] | • Acquired knowledge regarding requirements of the access network.<br>• Presented a dynamic access-aware bandwidth allocation method that follows every cell's dynamic traffic requirements.<br>• This step is followed by allocating the appropriate bandwidth from a shared spectrum pool.<br>• Presented ML-enabled method for efficiently predicting the network state and managing the available spectrum. | • The proposed approach is compared with two existing techniques considering spectrum allocation accuracy and utilization efficiency as the performance metrics. | • The considered performance metric is average utilization factor.<br>• The obtained results demonstrate that ML-enabled method can outperform existing techniques and achieve performances close to an ideal dynamic system. | • The article is limited to 5G networks scenario. |

from infrastructure providers. After listing out some of the benefits of utilizing blockchain for managing network, the authors proposed a blockchain-enabled algorithm for sharing unlicensed spectrum between network operators. Tokenization of spectrum is adopted to regulate the spectrum access and ensure the unselfish behavior of operators. Results from nu-merical simulations considering a two-operator scenario using a 5 GHz ISM band were presented. The authors showed how their proposed algorithm balances throughput and tokens with respect to time. Nevertheless, the paper lacks blockchain-based deployment with a multi-operator scenario.

In [102], the authors discussed blockchain usage for managing radio spectrum, particularly related application for dynamic SS. Interestingly, the authors presented various key blockchain technology characteristics such as decentralization, transparency, immutability, availability, and security for improving different aspects of SS. For instance, decentralization is useful for the elimination of third parties (e.g., spectrum licensees and band managers); transparency can enable a clear view of SS and real-time SS; immutability allows tamper-proof recording, which leads to audibility; and security can enable high reliability and less fraud in the SS process. Moreover, using smart contracts enables the establishment of complex SS rules in a transparent and trustless manner. To better understand the versatile use of blockchain technology for SS, the article focused on the four different categories of SS. These categories are primary cooperative, primary non-cooperative, secondary cooperative, and secondary non-cooperative sharing. As per the authors, it primarily implies all the users have an equal stand as far as spectrum access rights are concerned; however, secondary means there are hierarchical rights levels. On the other hand, cooperation implies pre-defined agreements exist between users, whereas non-cooperative means users do not follow any agreements. Various pros and cons of using blockchain for these four categories are also explained in detail.

The authors in [103] presented a summary of the standardization progress related to SS and blockchain application for the wireless networks. Thereafter, the authors investigated the critical issues encountered by the blockchain-enabled SS techniques, following which probable solutions presented. Lastly, directions for research related to blockchain enabled dynamic SS are presented.

In [104], the authors investigated the dynamic SS process in the B5G enhanced URLLC services. The authors merged the blockchain technology with a B5G hybrid cloud for realizing SS between the IoT devices. Further, the authors proposed a radio spectrum resource sharing structure in URLLC based on RL, which includes resource classification, resource scheduling, and neural network optimization. In the proposed design, blockchain and hybrid cloud are integrated with an aim of building a public management platform for registering and managing the IoT devices information using the unified coding and identification standards.

Authors in [105] leveraged blockchain technology and RL to propose a RAN slicing framework for B5G networks. This learning-based slicing framework is capable of managing SS. Specifically, blockchain technology improves transparency and fairness among multiple parties involved and increases orchestration efficiency. The authors considered the case of base station having the responsibility for exercising dynamic SS. Moreover, a multi-agent system model, where agent is cognitive user and environment comprises of all primary and secondary users in network, is considered. Finally, the authors regard the scenario of 5G and B5G networks (where a proliferating number of users request limited spectrum resources) as a problem of learning in a multi agent system. The use of private blockchains such as hyperledger fabric is discussed in the study. The authors conducted simulations to show that their proposed framework is energy efficient and provides fast convergence.

The authors in [106] presented a consortium blockchain dependent DSS framework for enabling efficient and secure DSS which guarantees QoS and revenue for 5G networks. In proposed framework, regulators can supervise the complete DSS process thereby guaranteeing every participant's revenue. Further, every MNO on chain is able to adaptively function as provider of spectrum or as requestor of spectrum depending on the demand. The recording of spectrum allocation is over chain through smart contract. Further, solution to optimal pricing of spectrum and strategies for purchase are obtained via a multi-leader multi-follower Stackelberg game model. Next, solutions to equilibrium are obtained using proposed algorithm. Thereafter, the authors have built a prototype using the Hyperledger Fabric consortium blockchain and mean delay has been evaluated. The feasibility of the proposed blockchain based DSS method is validated and it is demonstrated that the mean delay increases as participants number grows.

In [107], the authors briefly introduced the blockchain fundamentals followed by a comprehensive investigation of recent efforts for including blockchain within wireless networks. As a key contribution, the authors proposed blockchain radio access network with unified framework. The network uses blockchain technology with high security and efficiency. The key parameters such as smart contract, mathematical modeling, cross network sharing, and data tracking and auditing were elaborated. Further, a prototype design of proposed framework was detailed. The results demonstrated advantages of blockchain radio access network in terms of lesser service delay, high useage of resource, and important balance of load.

The authors in [108] investigated blockchain and SDN empowered dynamic spectrum management. Therein, the operators obtain incentives by sharing their resources to a common resource pool. Initially, the authors proposed a blockchain enabled framework to manage spectrum considering an integrated network involving inter slice SS and intra slice allocation of spectrum. In specific, inter slice SS depends on blockchain consortium generated by upper-tier SDN controllers, whereas intra slice spectrum assignment is based on a graph coloring enabled channel assignment algorithm. A bilateral confirmation protocol and a consensus mechanism were also proposed for consortium blockchain. The results demonstrated that for reaching a consensus, the proposed algorithm incurs lesser time compared to the practical Byzantine fault tolerance algorithm. Further, proposed channel allocation method was shown to significantly improve spectrum usage and outperform existing method.

In [109], the authors proposed a Blockchain based multi operator service provisioning for 5G users with intra- and inter-spectrum management amongst the numerous telecom operators. In specific, the authors presented a Blockchain enabled model for SS between operators with an aim of minimizing the spectrum under-utilization. The delay aware operations of proposed blockchain framework were assessed and the results demonstrated the proposal's scalability, as it can simultaneously provide transparency, security, and efficient SS operations.

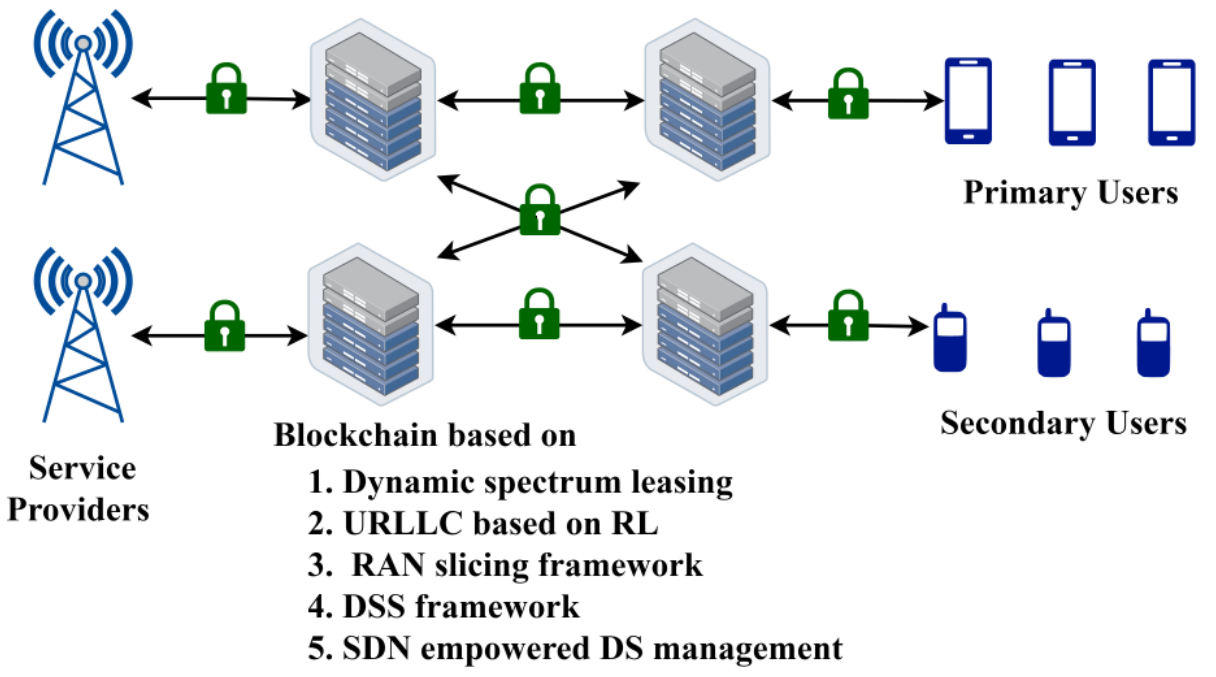

Fig. 6. Blockchain technology used for SS.

The authors in [110] proposed a Blockchain enabled decentralized platform for computing with trust. The platform provides self protection against cyberattacks on spectrum. The proposal introduced a real time secure SS technique for 5G networks in view of defending spectrum resources and applications. In addition, the proposal provides a decentralized and trusted platform for enabling secure SS and sharing of information between 5G spectrum. The results demonstrate that the proposed platform effectively bridges gap between supply of spectrum and demands in 5G networks thereby minimizing cost. The proposed method also ensures that (i) each transaction of spectrum is saved in block with security and reliability, and (ii) no payment is required for a third party platform. In addition, proposed method ensures that spectrum which is freely available is shared to save operator's cost and simultaneously improve spectrum usage.

*2) Summary*

To summarize, the existing centralized architecture to improve spectrum usage is inefficient as it is non-transparent and is highly vulnerable to multiple attacks. Therefore, blockchain enabled dynamic SS framework is required to implement a distributed architecture. This will provide advantages including decentralization, openness, transparency, immutability, and auditability. Further, introduction of blockchain technology, as shown in Figure. 6, within the SS technique also addresses multiple privacy issues and enables network to monitor and manage the spectrum usage and sharing efficiently. Overall, Blockchain-enabled SS will ensure decentralization, openness, transparency, immutability, and auditability. Table VI summarizes all the recent related studies that have implemented SS using Blockchain technology.

### D. *SS using other Techniques*

One can find in the literature several other techniques to implement SS in wireless networks, as shown in Figure. 7.

*1) Related Works*

In [111], the authors presented a two tier and multi tier approach for SS in which a small cell is considered for one tier, and a macro cell is considered for the backhaul network. Active and silent modes are used to partition the spectrum between these tiers. The silent mode works with the small cells for communicating and sharing the load, while the backhaul network does not operate. For the multi-tier network, two approaches have been discussed viz., (i) the first approach based on an analytical method uses the probability distributions for sharing, and (ii) the second approach finds all the combinations of User Equipment-Access Point associations in the 5G heterogeneous networks, followed by a frequency allocation. A centralized approach for maximizing the common achievable rate per small-cell base station is used to optimize the (i) access allocation or the backhaul channel, (ii) given channel to the transmission link. It is demonstrated that small cell base station(s) can have (i) an alloted channel for the access link only if a backhaul channel(s) has been allocated, and (ii) atmost single channel for an access link transmission. The results are presented as an analysis of backhaul channels numbers alloted to small cell base stations with minimum received signal power (RSP) and maximum RSP is presented. It is further shown that as RSP decreases, number of allocated channels decreases. The main limitations of study are that the proposed method (i) incurs a large amount of time and computational cost for operation, (ii) demonstrates limited usage of the backhaul network, and (iii) is not practically possible with both small-cells and backhaul network.

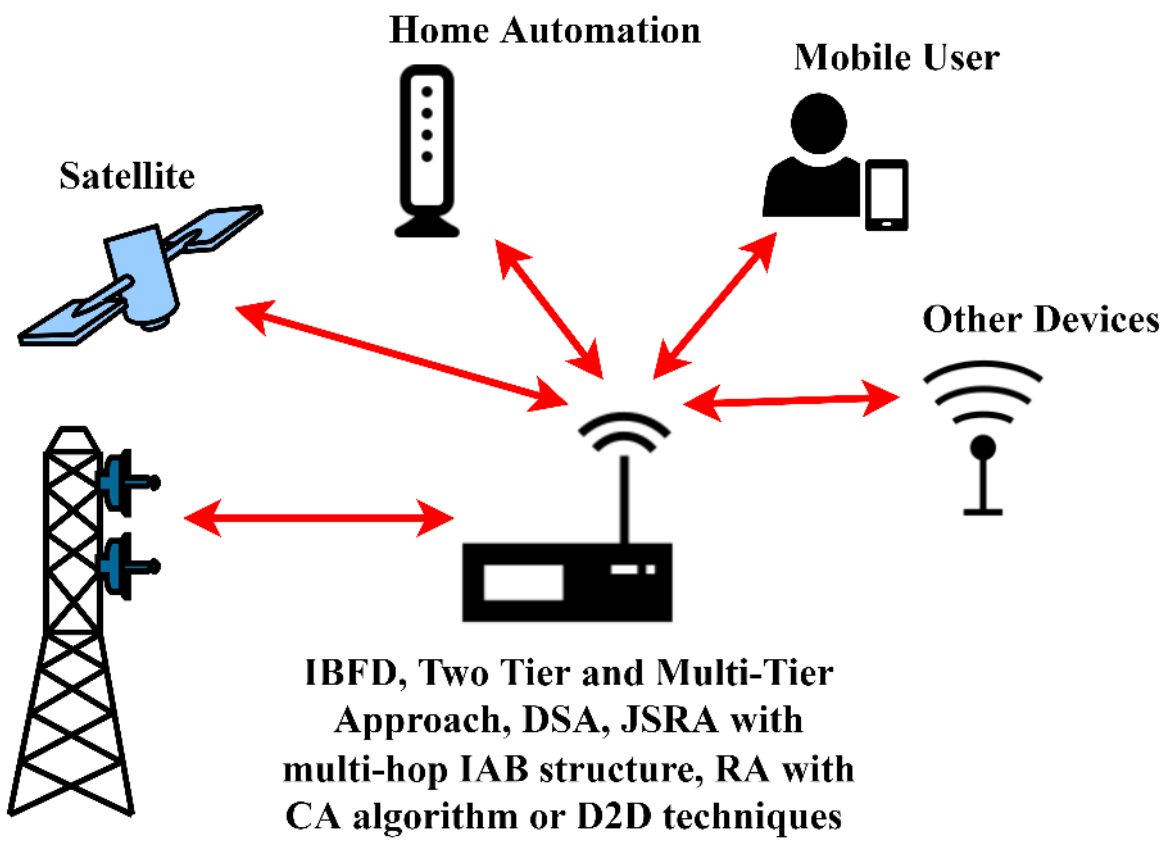

Fig. 7. Other techniques used for SS

The author in [112] formulated a Dynamic Spectrum Assignment (DSA) method for allocating spectrum in licensed experimental band 95 GHz to the Mobile Service Providers (MSPs) of a nation to attain high spectrum utilization. A comparison between number of customers of MSP and all leftover MSPs for set term of renewal (ToR) if first conducted, and next spectrum is assigned accordingly. The proposed DSA technique updates allotted spectrum to MSP during every ToR considering the variation in number of subscribers according to earlier ToR. The results showed that DSA outperforms the existing SSA (Spread Spectrum access), and that there is an optimization in the amount of required spectrum by every MSP during each ToR, and DSA method is useful in preventing the

TABLE VI
SUMMARY OF SS TECHNIQUES USING BLOCKCHAIN TECHNOLOGY.

| Reference | Aim | Methodology | Key Results | Limitations |
|---|---|---|---|---|
| [101] | • Convergence of blockchain and AI for 5G/B5G networks, and related benefits are discussed. • Proposed a blockchain enabled algorithm for generous sharing of unlicensed spectrum between network operators. | • Blockchain-based approach with spectrum tokenization to ensure unselfish behaviour during SS is presented. | • Simulation-based experiments are conducted to test throughput and account (token) balance. • The metric evaluated is divergence of throughput. | • Blockchain-based implementation is not presented. |
| [102] | • Detailed blockchain technology implementation for SS. • Highlighted evolution of SS methods in regard to satisfying different requirements to attain efficient spectrum utilization. | • SS technique has become the basis for dynamic and licensed shared access. • Explored application of blockchain to management of radio spectrum. | • Considered advantages and limitations of blockchain solutions. • Examined potential application of blockchain technology to four key SS categories. • The metrics considered are efficiency, cost, feasibility, and maintainability | • Assessment of how consistent the blockchain ideology is with SS is not investigated. |
| [103] | • Detailed standardization progress in regard to implementation of blockchain for SS technique. • Investigated the key issues which are encountered by blockchain-enabled SS techniques. | • Proposed potential solutions in regard to implementation of blockchain technology for SS technique. | • Identified future research directions in regard to blockchain-enabled dynamic SS. • The metrics evaluated are history of data inquiry, management of node, stability, auditability and privacy. | • Application of blockchain technology for DSS is not investigated |
| [104] | • Investigated dynamic SS process in the B5G enhanced eURLLC services. • Proposed a radio spectrum resource sharing structure in eURLLC based on RL which includes resource classification, resource scheduling, overall model, and neural network optimization. | • Merged blockchain technology with a B5G hybrid cloud for realizing SS between the IoT devices. | • Integrated blockchain with hybrid cloud can build a public management platform for registering and managing information of IoT devices using unified coding and identification standards. | - |
| [105] | • Proposed a RAN slicing framework for 5G and B5G networks which can efficiently handle SS. | • The framework uses blockchain technology to improve transparency and fairness among multiple parties and RL to mitigate issues related to dynamic spectrum access in 5G-core. | • Proposed a RAN slicing technique using a consensus algorithm and learning strategy for SS expansion in 5G-CORE. • The metrics evaluated are throughput, channel switching, and EE. | • Blockchain-based implementation and related results are not discussed. |
| [106] | • Presented a consortium blockchain based DSS framework for enabling an efficient secure and efficient DSS with guaranteed revenue and QoS for future wireless networks. | • Regulators supervise the complete DSS process to guarantee every participant's revenue. • Recording of spectrum allocation over chain through smart contract. • Solutions to optimal pricing for spectrum and strategies for buying are obtained via multi-leader multi-follower Stackelberg game model. • Solutions to equilibrium are obtained using proposed algorithm. • Prototype is built using Hyperledger Fabric consortium blockchain. | • Simulations and prototype evaluations validate feasibility of proposed blockchain based DSS method • Average latency is considered as the metric and increases with the participants | • It is not specified if the model is general and can be used for single-operator efficiently. |
| [107] | • Proposed unified framework of blockchain RAN | • Enhanced efficiency and security using blockchain technologies. | • Results demonstrated benefits of latency to be low in terms of service, utilization in terms of resource to be high, and prominent load balance for Blockchain RAN | • It needs to be investigated if increasing number of subnetworks accommodated by blockchain radio access can improve throughput performance |
| [108] | • Investigated dynamic spectrum management via the use of blockchain and SDN | • Blockchain enabled spectrum management framework considering integrated network is proposed. • Bilateral confirmation protocol and consensus mechanism for consortium blockchain is proposed. | • For reaching a consensus, proposed algorithm incurs lesser time compared to practical Byzantine fault tolerance algorithm. • Results prove that proposed channel assignment algorithm outperforms baseline algorithm in terms of improved spectrum utilization in simulation and as well in real world scenarios. | • The proposed blockchain technique is not evaluated with real blockchain platform. |
| [109] | • Blockchain enabled implementation model for SS between operators to minimize spectrum under utilization is presented. | • Model to implement Blockchain based multi-operator service provisioning for 5G users with intra- and inter-spectrum management amongst multiple telecom operators is detailed. | • Latency aware operations of the proposed blockchain framework is evaluated. • Proposed blockchain based scheme for 5G is scalable simultaneously providing transparency, security and efficient SS operations | • Investigation considering real time blockchain model is not presented |
| [110] | • Proposed a trusted computing platform enabled via decentralized Blockchain method | • Introduced a real-time secure SS technique for 5G networks | • Proposed platform effectively bridges gap between supply of spectrum and demand in 5G networks thereby minimizing cost. • Proposed method ensures safety of each transaction of spectrum and it does not need any payment to third party platform as it is saved in block in a reliable manner. • Proposed method improves overall spectrum usage by sharing free spectrum thereby efficiently saving operator's cost. | • Real time application is not detailed. |

payment case of customers and MSPs for spectrum that is not used.

The authors in [113] analysed existing spectrum management and valuation methods, and identified the major elements for consideration while spectrum value is defined and assessed, specifically considering B5G networks. The authors characterized elements of spectrum valuation for new 5G networks from different stakeholder roles. A case study on valuation of spectrum considering recent decisions on 5G spectrum was presented. The authors highlighted various ways to valuate spectrum for5G networks considering that their deployment is specific to location for complementing past generations which addressed a wide area coverage. Lastly, the authors identified that research is required to analyze decision on 5G spectrum around the globe with more availability of data, especially with migration to high frequency bands.

The authors in [114] described SS techniques in the in-band full duplex mode (IBFD) wherein spectrum is split into two parts between the backhaul and small cell for uploading and downloading the data, respectively. Hence, IBFD can support the operation of small cells in the FD mode and is also able to communicate between backhaul network and the FD modes. Specifically, two tier network comprising of IBFD based Pico-Base Stations (P-BSs) and Macro-Base Stations (M-BSs) is proposed as a novel approach by the authors. The authors analysed system performance in the DnL while M-BSs wirelessly backhaul the P-BSs. As the IBFD-enabled P-BSs are present, they use same frequency for backhauling themselves on DnL and UpL with M-BS. The MBSs being traditional non-IBFD stations, are required to divide spectrum between access and backhaul links. The authors also introduced an innovative HetNet architecture with controlling IBFD capability and derived the average rate and probability of coverage for a distinctive user in such a network. The precise coverage and parameters of rate for proposed IBFD HetNet are then achieved using apt mathematical derivations. The derived equations can be tracked and computable coverage expressions are used to analyze future IBFD-enabled HetNets systems via modeling effective SNR distribution. The main limitations of this study are that (i) further division of bandwidth between the backhaul network and the access communications results in coverage area of the spectrum being small, and (ii) there is leakage of signal from transmitter to receiver, which introduces interference.

In [115], the authors investigated SS problem for D2D communications considering ultra dense 5G networks. The authors proposed a technique for D2D traffic to work simultaneously with existing traffic. A two step distributed algorithm is developed to select mode and allocate channel (i.e., D2D mode or cellular mode) for mobile station (MS) to minimize interfering D2D links and capacity maximization. The scheme is implemented in a multi-cell scenario with the BS enabled by mm-wave. The implementation shows that capacity of access for both D2D and cellular traffic remains unaffected after SS; however, managing interference with an unknown number of users in D2D communication is difficult.

The authors in [116] presented a Resource Allocation (RA) optimization problem with CA. An iterative allocation CA enabled decentralized algorithm is presented for optimally allocating resources of both primary and secondary carriers among the users. The authors also introduced a price selective algorithm for optimally assigning resources from various carriers among users. Minimum QoS is provided to all users, and applications were assigned higher priority through the proposed algorithms. The algorithm convergence with varied carriers was analyzed, and optimal convergence of algorithms was shown via simulations. The study is limited in the fact that developing RA with a CA framework for heterogeneous radio access technologies is challenging.

In [117], to enhance performance, some algorithms are proposed for enabling flexible assignment of DnL and UpL transmission while allocating and supporting both half- and full-duplex modes. Specifically, proposed algorithms can fulfill multiple requirements of performance for point to point and point to multipoint communications. The authors implemented binary interference classification for designing scheduling algorithms. Further, for obtaining feasible solutions, problem is systematically decomposed into three sub-problems: transmission power allocation, simultaneous scheduling of transmission, and transmission duration allocation. Depending on this, for obtaining solutions to optimization problem depending on static decisions of routing, following heuristic scheduling algorithms were proposed: (i) conflict graph maximum independent set (CG-MIS) algorithm, (ii) water-filling power allocation algorithm, and (iii) proportional fair transmission duration algorithm. Also, a dynamic routing (DR) algorithm was proposed as static decisions of routing may degrade system performance when multiple users traffic shows congestion at a few nodes of network. Extensive simulations were conducted considering multiple system parameters to evaluate system performance and to analyze the proposed algorithms considering different frame structures. Potentialities of proposed algorithms in enhancing system performance are illustrated. It was also shown that proposed algorithms attain major gain over existing schemes in terms of low latency and data rate and closely approach theoretical optimum. As a scope for future research, solutions for issues in multi-point communication in HetNets need to be implemented.

The authors in [118] proposed a technique, namely, SwarmShare, which provides a mobility-resilient SS framework in view of networking the unmanned aerial vehicles in the 6 GHz band. Initially, the authors formulated the mathematical framework of the SwarmShare problem to maximize SE of unmanned aerial vehicles network. The formulation can jointly control flight and transmission power of unmanned aerial vehicles, simultaneously associating them with terrestrial users under the consideration of interference from the primary users. The authors found that no closed form mathematical models can characterize statistical behaviors of total interference from unmanned aerial vehicles to primary users. Subsequently, the authors proposed a data-driven 3-phase SS technique that includes starting power enforcement, offline dataset guided online power adaptation, and RL-based unmanned aerial vehicle optimization. The obtained results demonstrated that total interference from unmanned aerial vehicles to primary users can be effectively controlled below target level using the SwarmShare method without needing real time cross system channel state information.

In [119], the authors modelled a non-orthogonal SS system considering large network in which many seller mobile network operators rent the licensed sub bands to multiple buyer mobile network operators. Initially, the authors analyzed the expected rate, and mobile network operator expected profit for every user through stochastic geometry. Next, they formulated joint problem of controlling power and sharing licensed sub band for maximizing envisioned gains of all the MNOs as multi-objective optimization problem considering users' QoS need and non-negative return on investment constraints. The authors used a combination of the $\in$ -constraint and weighted sum methods for transforming multi-objective optimization problem within one objective form. However, it is found that the transformed problem is non-convex owing to binary variables and non-convex rate functions in objective function and constraints. The authors used a penalty function and approximated non-convex rate functions via method enabled by constrained stochastic successive convex approximation.

Lastly, results demonstrated correctness and performance of proposed algorithm considering multiple conditions.

The authors in [120] investigated the intelligent reflecting surface-enabled SS method considering model for interference, efficiency in estimating channel, and beamforming design which is passive and robust. The authors characterized interference in SS system comprising of one pair of primary user and one pair of secondary user, and extended the same to large network by leveraging the Poisson point process. Next, the authors proposed an efficient framework to estimate channel. In addition, SE and EE trade-off was analyzed in view of the channel estimation accuracy. The authors also discussed design of beamforming which is passive and robust considering estimate of channel which is imperfect and non-ideal shifts of phase.

#### 2) Summary

To summarize, multiple SS techniques have been introduced in literature to increase SE. These SS techniques have varied features and can be evaluated considering various parameters. Also, they differ in the operation mode and management of interference, including cases where SS needs to be implemented. The majority of SS methods can be categorized in accordance with the co-existence or co-operation, and sharing among the equals or primary-secondary sharing. Specifically, when a multitier approach for SS is implemented, it must be ensured that allocation of backhaul channels is optimized. It has been shown that small-cell base stations with assigned backhaul channel(s) consist of assigned channel for the access link. Further, in such a scenario, each small-cell base station has atmost one channel for an access link transmission. IBFD can be implemented to support the operation of small-cells in FD mode thereby ensuring communication between the backhaul network and the FD modes. It has been shown that mean rate for P-BS-associated user can be modeled as minimum of rates on link pair. To implement SS for D2D communications, the interfering D2D links can be reduced and the capacity can be maximized. It has been shown that access capacity for D2D and cellular traffic is not affected after sharing the spectrum. Further, allocating resources using CA is also possible for solving the various utility functions. It has been shown that such methods are robust and converge to optimal rates irrespective of available resources being excess or in shortage. Also, the SS technique can be implemented using non-orthogonal methods which have been shown to provide correct and high performance under various conditions. Further, intelligent reflecting surface-enabled SS methods can also be implemented to model interference, provide efficient channel estimation and design robust passive beamforming.

Overall, factors such as SE gains, protection from interference, network congestion, and mobility support, must be accounted for when choosing a SS technique. Also, no 'best' SS technique exists as it largely depends on the application for which it is implemented. Table VII summarizes all the recent related studies that have implemented SS using multiple technologies.

### E. Challenges

Previous sub-sections have clarified that inefficient spectrum utilization will be a major bottleneck in intelligent wireless networks. Hence, instead of exclusively allotting spectrum, most obvious solution is sharing the available spectrum. However, SS introduces multiple challenges which, if not efficiently addressed, will result in degrading the co-existing networks' performance levels. Various challenges exist in the implementation of SS for intelligent wireless networks. The most significant challenges include the following:

1) Accurate implementations and related analysis for integrating NOMA with CR are required. This demands proper formulations of advanced SS techniques to use the available spectrum efficiently. These techniques must address the issues of information location, privacy and security, dynamic radio range, and spectrum sensing. Also, accounting for the primary users' saturated and unsaturated sources and traffic to further improve the system throughput requires research attention.
2) Enabling intelligence within wireless networks mandates the use of advanced ML mechanisms and model formulations to (i) counter high cell-to-cell interference, (ii) provide system feasibility to operate in outdoor scenarios, and (iii) ease the connection between two cells regarding bandwidth sharing. Also, for such networks, availability of information must be continuously ensured, which requires advanced techniques that must account for energy limitations, and must ensure real-time operations and fast alteration in the operation environment. Further, dynamic processes must be implemented when implementing game models so that information is always available, and when implementing the MDP models, a stationary environment must be ensured.
3) Blockchain enabled SS will present the challenge of developing a lightweight blockchain solution for low cost IoT devices and using high performance decentralization techniques for B5G. Also, blockchain techniques will be required to account for aspects related to security and privacy, and protocols applicable to wireless channel. Finally, fundamental limits pertaining to performance and security evaluation will also have to be developed.
4) There is a fundamental need to develop a framework combining RA with CA under the consideration of spectrum access differences and physical layer of considered technologies.
5) A major challenge will be to develop mechanisms for RA with CA framework for heterogeneous radio access technologies. This needs algorithms enabled by CA for managing resource where user equipments connect to an eNodeB and access points simultaneously being allocated with resources based on CA. Such an approach is challenging because technologies will have differences in access of spectrum and physical layer.
6) The development of efficient component carriers' assignment and algorithms for the networks incorporating LTE-A with CA will be required. It must be ensured that component carriers involved in assignment process

TABLE VII
SUMMARY OF SS TECHNIQUES USING MULTIPLE OTHER TECHNOLOGIES.

| Reference | Aim | Methodology | Key Results | Limitations |
|---|---|---|---|---|
| [114] | • Analyzed the FD technique.<br>• Described regarding SS in IBFD. | • IBFD-enabled P-BSs and M-BSs for a two tier network is employed. | • Coverage probability and covered rate are used as the performance metrics.<br>• It is shown that IBFD supports and communicates between backhaul network and FD modes. | • Coverage area of spectrum is small and interference is introduced. This aspect needs investigation. |
| [111] | • Tier-based SS, two-tier and multi-tier approach for SS are detailed. | • Distributed backhaul channel allocation and allocation of more backhaul spectrum for the cell-edge SBSs is presented. | • The number of allocated backhaul channels and coverage probability are used as metrics.<br>• Comparative analysis of Minimum and Maximum RSP is conducted.<br>• It is found that low interference permits a high number of cell-edge SBSs to obtain much better rates provided sufficient backhaul coverage exists. | • Proposed method involves a large amount of time and computational cost.<br>• It is limited in usage of the backhaul network. Proposed approaches are practically not possible with both small cells and backhaul networks. |
| [112] | • Fformulated DSA for spectrum allocation. | • Experiment on 95 THz band has been conducted using DSA technique. | • The performance metrics considered in the study include mean capacity, SE, EE, and cost efficiency.<br>• Results prove that users need not pay if they are not using spectrum. | • Only small cell scenario is considered for a large cell environment. |
| [115] | • Discussed associated SS sharing problem in ultra dense 5G cellular networks for D2D communications.<br>• Proposed simple method for D2D traffic work simultaneously with cellular traffic. | • A two step distributed algorithm is developed to select mode and allocate channel for MS to minimize interfering D2D links and capacity maximization.<br>• The scheme is implemented in a multi-cell scenario with a base station enabled by mm-Wave. | • The number of successful flow rate traffic is considered as the metrics.<br>• Access capacity for both D2D and cellular traffic is unaffected after SS. | • Managing interference with an unknown number of users in D2D communication is difficult. |
| [113] | • In the context of 5G networks, authors aimed to characterize identified spectrum valuation elements in consideration with different stake holders.<br>• Introduced spectrum valuation with the aid of case study presenting recent 5G spectrum decisions in 3.5GHz. | • The metrics considered include offered services, location-specific characteristics, and band-specific characteristics.<br>• With spectrum valuation perspective, 5G spectrum awarding results are accessed. | • Case study on 3.5 GHz for six European countries is conducted, and spectrum valuation methods are shown. | • Research is needed on analyzing 5G spectrum decisions globally as more data becomes available. |
| [117] | • Efficiently solved the maximization problem.<br>• Formulated optimization problem in HetNet with multihop IAB structure for joint scheduling and resource allocation. | • The performance metrics considered include latency, data rate, and throughput.<br>• Binary interference classification, which prevents two links from being active at a time, is implemented. | • It is shown that flexible adjustment of DnL and UpL transmission duration and allocation is possible.<br>• Fulfilling different performance requirements for both point-to-point and point-to-multipoint communications is also possible. | • The system performance may degrade due to fixed routing decision. The proposed algorithm has addressed this problem. |
| [116] | • Introduced CA algorithm for iterative decentralized rate allocation for primary and secondary carriers' resources.<br>• Optimally allocated multiple carriers among users using CA algorithm in a price-selective centralized RA. | • Novel Price selective centralized algorithm is proposed. | • The metrics used are data rate and utility function.<br>• The algorithm's robustness, which converges to optimal rates in case of scarcity of resources, has been shown. | • Developed RA with CA framework for heterogeneous radio access technologies such as LTE-A and WiFi.<br>• Such an approach is challenging because the two technologies have differences in spectrum access and physical layer. |
| [118] | • Jointly controlled the flight and transmission power of unmanned aerial vehicles.<br>• Simultaneously associated them with terrestrial users under the consideration of interference from primary users.<br>• SwarmShare technique is proposed, which provides a mobility-resilient SS framework in view of networking the unmanned aerial vehicles in the 6 GHz band.<br>• Proposed a data-driven 3-phase SS technique that includes initial power enforcement, offline dataset-guided online power adaptation, and RL-based unmanned aerial vehicle optimization. | • Formulated mathematical framework of the SwarmShare problem to maximize SE of the unmanned aerial vehicles network. | • The considered performance metrics include average throughput, outage probability, and capacity.<br>• It is shown that no closed-form mathematical models can characterize statistical behaviors of aggregate interference from unmanned aerial vehicles to primary users.<br>• Total interference from unmanned aerial vehicles to primary users can be effectively controlled below target level by using the SwarmShare method without any need for real-time cross-system channel state information. | • The proposed techniques consider use-case of unmanned aerial vehicles. Other use cases are not discussed. |
| [119] | • Formulated joint problem of controlling power and sharing licensed sub-band to maximize envisioned gains of all MNOs as a multi-objective optimization problem considering users' QoS need and non-negative return on investment constraints.<br>• Modelled a non-orthogonal SS system considering a large network in which many seller MNOs rent licensed sub-bands to many buyer MNOss.<br>• Used a combination of $\in$ -constraint and weighted sum methods for transforming multi-objective optimization problem to a single objective form. | • Analyzed the expected rate and MNO's expected profit for every user through stochastic geometry.<br>• Transformed multi-objective optimization problem within a single objective form using a combination of $\in$ -constraint and method of weighted sum. | • Average profit of buyer mobile network operators and expected rate of users are the considered metrics.<br>• It is shown that transformed problem is non-convex since objective function and constraints consist of binary variables and non-convex rate functions.<br>• Correctness and performance of proposed algorithm is observed considering different conditions | • The case of licensed users is considered and the MNOs are considered to own the spectrum. Other cases are not considered. |
| [120] | • Investigated the intelligent reflecting surface-enabled SS method from interference modeling perspective, efficient estimate of channel, and beamforming design which is passive and robust.<br>• Characterized interference in a SS system consisting of one pair of primary user and one pair of secondary user.<br>• The above is extended to large network by leveraging the Poisson point process.<br>• Efficient framework for estimating channel is proposed. | • Analyzed trade-off between SE and EE in view of channel estimation accuracy.<br>• Discussed design of beamforming in presence of imperfect channel estimation and non-ideal shifts of phase. | • The considered performance metric includes achievable rate of secondary user.<br>• Results show the effectiveness of employing intelligent reflecting surfaces to improve the secondary user rate.<br>• Results also show the advantages of encountering challenging interference cases in existing SS systems without the intelligent reflecting surface. | • This is only an initial study, further investigations are required. |

of resource should not be assigned based on RSP level, estimated by each user equipment, as it is insufficient in the case of CA, and this is a crucial challenge. Formulating methods to mitigate interference caused by CA must be considered. This can be ensured by choosing an optimal group of component carriers for CA in an LTE-A network.

7) Investigation must be conducted for jointly optimizing the link scheduling, RA, and routing strategies to improve overall performance of system compared to existing solutions, considering an increase in computational complexity.

8) THz frequency provides high data rates; however, it must solve a key issue of transmission of data for longer distances owing to higher loss of propagation and atmospheric absorption [114]. A novel transceiver design architecture is required for THz systems to ensure complete use of the wide available bandwidths. Also, less gain and effective area of antennas operating at THZ frequency is another challenge. Lastly, the health and safety concerns must be addressed.
9) The migration of M-MIMO from the 5G system to the B5G networks needs advanced study on developing effective architecture. Here, challenge is to integrate all systems within one platform. Management of beam which is efficient is a challenge for M-MIMO systems. It will also be important to select optimal beam efficiently in advanced applications such as high-speed vehicular systems.
10) After division, bandwidth between access communications and backhaul network, which results in a coverage area of the spectrum, is small. This will require investigation of efficient bandwidth allocation strategies. Further, solutions for multi-point communication in the HetNets which will further enhance communication need much research attention.
11) Design of beamforming which is passive and robust in presence of imperfect channel estimation and non-ideal shifts of phase must be investigated to make SS process effective. Also, the signal leakage from transmitter to receiver leads to interference and presents a major challenge.
12) For any technique implemented to advance SS for intelligent wireless networks, it will be required to account for the execution time and computational costs.

Overall, from the above, it is clear that the SS techniques implemented in the 5G wireless networks cannot be extended to the B5G wireless networks. Instead, novel and advanced SS and related methods will be required in the B5G wireless networks. Towards this end, only a few studies have formulated and implemented the advanced SS techniques for intelligent B5G wireless networks. In next section, we present a summary of such studies that have focused on implementing SS techniques for B5G networks.

## IV. SS Techniques for B5G Wireless Networks

In B5G wireless networks there will be significant flexibility for spectrum occupancy which will open the interfaces introduced through O-RAN with the addition of RAN Intelligent Controller (RIC). As a result, spectrum can be managed efficiently, and emerging open interface standards can be explored for minimizing interference between heterogeneous systems in a very congested B5G radio environment. Also, O-RAN will permit operations in different spectrum bands, involving varied SS levels over the mm-wave and THz spectrum. Hence, it will be required to design a system architecture including SS methodology enabled via O-RAN RIC.

Currently, there are major access issues between the spectrum band users, and this will grow further in the B5G wireless networks which will have to deal with dynamically changing environments. Hence, existing SS techniques will not suffice and advanced methods with new rules, conditions and mechanisms are required. Further, primary users will need to be protected from interference due to unlicensed spectrum users, whereas new users will require equal spectrum access opportunities. Hence, a database-enabled model will be a need of the hour to resolve the aforementioned issues. Lastly, operations within different spectrum bands, which will range between mm-waves to THz and between the licensed and the unlicensed spectrum band, will require sharing frequencies through advanced CA techniques.

### *A. Related Works*

The exponentially increasing demands will push the B5G networks to adopt SS techniques for improving the network's ability to monitor, access, use, and share spectrum. The implementation of SS in the B5G intelligent wireless networks will mandate novel designs, optimization methods, and processes for measurements, simultaneously with the development of newer standards and techniques. Further, such networks will have to cater to heterogeneous demands and will have to operate in licensed, unlicensed, license-assisted, or tiered access bands for which advanced algorithms and hardware must be developed.

In [5], authors proposed a B5G architecture as integrated system of enabling technology(s). Potential challenges in developing B5G technology were discussed, and possible solutions were proposed. Moreover, the 5G core services, namely URLLC, mMTC, and eMBB, were reviewed. The study concluded that in order to fulfill the service quality improvement demands, an increasing wireless services must be optimized for more than one requirement.

In [121], the authors provided insight into the mm-wave and THz applications including the security aspects, design challenges, role of AI, and key enabling technologies in the B5G wireless networks. The study described measurements, wave propagation, and channel modelling in the THz bands. Design of multi-user-ultra-M MIMO hybrid beam forming system for THz frequencies was also proposed with many independent data streams per user. Different Quadrature Amplitude Modulation (QAM) and antenna array configurations at transmitter and receiver were analysed. The results of multi-user-ultra-M MIMO hybrid beamforming system operating at mm-wave and THz were compared to demonstrate feasibility of using THz frequencies for the future wireless networks. The results suggested the use of a specific mMIMO antenna configuration depending on independent data streams amount for every user. It was also recommended to utilize larger data streams for every user for achieving higher throughput.

In [122], spectrum management approaches were proposed. The spectrum's engineering, economic, and strategic value was analyzed. Few challenges in the B5G networks such as equality, efficiency, and control were also discussed.

In [123], authors introduced many emerging technologies and applications for B5G such as space-air-ground integrated networks, industrial IoT, THz communication, and ML. Chan-

TABLE VIII
SUMMARY OF SS TECHNIQUES IN B5G INTELLIGENT WIRELESS NETWORKS.

| Reference | Aim | Methodology | Key Results | Limitations |
|---|---|---|---|---|
| [121] | • Provided insight into mm-Wave and THz applications.<br>• Explored design challenges, key enabling technologies, role of AI, and security aspects in B5G wireless networks.<br>• Described wave propagation, channel modelling, and measurements at THz bands.<br>• The possibility of using ultra-massive MIMO hybrid beam forming techniques for improved front-end performance at frequencies above 100 GHz is detailed. | • Parallel Decomposition technique, QAM Modulation to verify system performance at 28GHz,73 GHz, and 140 GHz frequencies is implemented. | • 3D radiation patterns for multiple BS antennas with 8 users in MU-massive MIMO hybrid beam forming systems are demonstrated.<br>• It is shown that Root Mean Square Error Vector Magnitude (RMS EVM) values are more for users with a smaller number of independent data streams, whereas RMS EVM values are minimum for users with more independent data streams.<br>• RMS EVM is used as the performance metric | • Designing hybrid beam forming for users with a less (or single) number of independent data streams by minimizing RMS EVM values and achieving higher throughput is still under research. |
| [122] | • Provided a comparative analysis of spectrum management approaches such as administrative allocation, market based mechanisms, and the unlicensed commons approach.<br>• Analysis of the spectrum's economic and strategic value is conducted. | • Discussed the role of SS in the B5G era and related enabling technologies.<br>• Study emphasized different aspects in B5G simultaneously accounting for challenges arising from the underlying regulatory environment. | • EE is the considered as the metric.<br>• Results on equality and EE issues in B5G networks are provided. | • Solutions for efficient energy management in B5G is still an open issue. |
| [123] | • Introduced several emerging technologies and applications for B5G.<br>• Review of measurements for channel and models for technology(s) and application(s) is presented. | • Channel research is conducted in view of designing B5G wireless communication systems | • Outage probability is proposed as the performance metric.<br>• Outlook for B5G channel measurements and models is discussed. | • IIoT communication scenarios are rich scattering because there are many metal objects which makes channel models for traditional cell communications unable to describe channel correctly. This aspects needs investigation. |
| [124] | • Highlighted most promising research from recent literature in common directions for B5G networks. | • Explored key issues and potential features of B5G communications | • The research topics are thoroughly examined for obtaining specific conclusions. | • The challenges associated with B5G networks stem from five key components which need further investigation. |
| [5] | • Provided a study on integrated system of enabling technologies.<br>• Illustrated an integrated system of the enabling technologies for four typical urban application scenarios. | • Potential challenges in development of B5G technology are discussed and possible solutions are proposed. | • Opportunities to explore B5G networks are analyzed. | • Only five B5G core services are identified, and eight key performance indices (KPIs) are detailed. |
| [125] | • Discussed emerging technologies.<br>• Review of Optical Wireless Communication (OWC) technologies to demonstrate their effectiveness for successful deployment of 5G/B5G and IoT systems is presented. | • Key characteristics of 5G and IoT networks are discussed.<br>• Possible B5G needs and various OWC technologies are briefly presented.<br>• The scope of OWC technologies to meet 5G/B5G and IoT requirements is detailed.<br>• Challenging issues related to OWC deployment for 5G/B5G and IoT solutions are discussed. | • Presented the desired services with needs and prospective technology(s) for B5G networks. | • Outlined possible challenges and research directions to obtain proposed goals of B5G; however, solutions are not highlighted. |
| [126] | • Addressed aspects related to integrated satellite-terrestrial communication networks (ISTCNs).<br>• Presented few open research issues and challenges in regard to SS in ISTCN.<br>• Highlighted that with terrestrial networks starting to utilize higher spectrum, these bands overlap with those used by satellite networks.<br>• It is shown that there is a potential in satellite terminals to demonstrate integrated access to the terrestrial networks. | • Performance of satellite-terrestrial SS process via the introduction of NOMA and CR improves when ISTCN attempts to access spectrum.<br>• The new technique, CR-NOMA, which integrates NOMA and CR, allows different satellite terminals to access both idle and busy spectrum simultaneously. | • Power and interference are the considered performance metrics.<br>• Proposed process results in high efficiency owing to complete access of the spectrum. | • The proposed technique will require management of spectrum for minimizing congestion and interference, which may result from the SS process. This aspect is not addressed in the study. |
| [127] | • Addressed SS issues in the Ubiquitous IoT (UIoT), a 3-D network spanning the space-air-ground.<br>• Proposed a strategy that migrates SS of the B5G system to a hybrid blockchain system. | • Implemented blockchain and B5G hybrid cloud techniques. | • SE is used as the performance metric.<br>• It is shown (i) that by sub-dividing spectrum granularity of UIoT devices within different dimensions, it is possible to securely and reliably ensure that a large amount of UIoT terminal data can be accessed, and (ii) massive data interconnection and externally perceived data authenticity can be ensured via cloud computing and smart contract technology. | • Performance analysis of proposed strategy is not performed in the article. |
| [128] | • Addressed UDN issue in B5G networks.<br>• Proposed a distributed Q-learning-enabled multi-dimensional SS security method for implementation in the mm-wave band. | • The proposed method uses a fine-grained SS security framework based on spatial multiplexing and can balance performance and security of the UDN.<br>• For modelling and analysis, a non-cooperative game theory approach is used to ensure multi-dimensional SS security and optimal spectrum allocation. | • The validity of proposed method is established via extensive simulation experiments considering access delay and throughput as the performance metrics.<br>• The results demonstrate that proposed method (i) minimizes user's access delay significantly, (ii) enhances accommodation of an increased number of unauthorized users, and (iii) improves network throughput simultaneously achieving high network security performance. | • The key research question which requires a solution is ensuring customers' safety while simultaneously provisioning universal services to multiple users. This aspect is not addressed in the study. |
| [129] | • Addressed the problem of supporting large wireless data traffic amount over a limited spectrum.<br>• Proposed a green SS solution for B5G network via the crowdsensing method to construct an accurate crowdsourced spectrum database cost-effectively. | • ML-based classifier is implemented for selecting such spots which are required to be interpolated by crowdsensing users.<br>• An offline and an online user incentive mechanism is proposed to attract customers to participate in crowdsensing task. | • SE is used as the performance metric.<br>• The proposed solution is validated via the conduction of a vehicle-based measurement campaign over a 2km×7km area considering real-time collected data. | • The validity of the proposed method for a large B5G network is not conducted in the article. |

nel measurements and models were reviewed for these technologies and applications. In the case of the THz channel, it was suggested that dependency of channel parameters on frequency needs must be considered in the models due to large frequency(s) range.

The authors in [124] highlighted significant studies with core contributions involving exploration of key issues and vital properties of B5G networks. The study provides a broad outline of research topics on B5G communications, covering recent industry developments in major service areas and corre-sponding difficulties. The study also highlights trade-offs set, including major challenges and prospective solutions for B5G networks.

The authors in [125] envisioned future B5G wireless communications and network architectures. Trending technologies which can assist B5G architecture development in guaranteeing QoS were presented. These technologies include THz communications, AI, blockchain, unmanned aerial vehicle (UAV), 3-D networking, quantum communications, cell-free communications, and holographic beamforming. Expected applications and corresponding requirements together with potential challenges and research directions were also outlined.

The authors in [126] addressed the aspects of integrated satellite-terrestrial communication networks (ISTCNs), which have emerged as a key area for research. The authors highlighted that with the terrestrial networks starting to utilize higher spectrum, there is now occurring an overlap of the bands with those used by the satellite networks. Hence, there is a potential in the satellite terminals to provide integrated access to the terrestrial networks. In turn, this will require management of the spectrum for minimizing congestion and interference, which may result due to the SS process. Hence, the authors proposed a satellite-terrestrial SS process via the introduction of NOMA and CR when ISTCN attempts to access the spectrum. The new technique, CR-NOMA, which integrates NOMA and CR, allows different satellite terminals to simultaneously access both, idle and busy spectrum. Such a process, in turn, results in high efficiency owing to a complete access to the spectrum. Lastly, the authors presented some few issues and challenges for research on SS in ISTCN.

In [127], the authors addressed SS issues in Ubiquitous IoT (UIoT), which is 3-D network spanning space-air-ground. The authors stated that due to the requirements of sharing amongst the various network operators, it is mandatory to advance research in the domain of SS for B5G wireless networks, which can be attained via the implementation of blockchain and B5G hybrid cloud techniques. The authors proposed a strategy for migrating SS of the B5G system to a hybrid blockchain system. It is shown that (i) by subdividing spectrum granularity of UIoT devices within different dimensions, it is possible to securely and reliably ensure that a large amount of UIoT terminal data can be accessed, and (ii) massive data interconnection and externally perceived data authenticity could be ensured via cloud computing and smart contract technology.

The authors in [128] stated that the ultra-dense network (UDN) issue in B5G networks has recently received much attention. The key research question requiring solution is that of ensuring safety of the customers simultaneously provisioning universal services to multiple users. The authors proposed a distributed Q-learning-enabled multi-dimensional SS security method for implementation in mm-wave band. In the proposed method, a fine-grained SS security framework was used, which is based on spatial multiplexing and can balance the performance and security of UDN. The validity of proposed method is established via simulation experiments. The results demonstrate that proposed method can (i) minimize user's access delay significantly, (ii) enhance the accommodation of an increased number of unauthorized users, and (iii) improve the network throughput while simultaneously achieving high network security performance. Lastly, the authors also a designed model based on non-cooperative game theory approach. The analysis of this model demonstrated that it can ensure multi-dimensional SS security and optimal spectrum allocation.

In [129], the authors addressed the problem of supporting large wireless data traffic amount over a limited spectrum. The authors proposed a green SS solution for B5G network via the crowdsensing method to construct an accurate crowd sourced spectrum database cost-effectively. Specifically, an ML-based classifier was implemented for selecting such spots that are required to be interpolated by the crowdsensing users. Further, an offline and an online user incentive mechanism was proposed to attract customers to participate in the crowdsensing task. The proposed solution was validated via the conduction of a vehicle-based measurement campaign over a 2km $\times$ 7km area considering real-time collected data.

### *B. Summary*

To summarize, the B5G wireless networks will involve small-cell deployments, which will result in high geographical and traffic density, thereby requiring SS between the incumbents and the unlicensed users. Specifically, several studies have described the measurements, wave propagation, and channel modelling at the THz bands. It has been shown that for THz channel, dependency of channel parameters on frequency needs must be considered in models due to extensive frequency(s) range. Also, the B5G architecture has been proposed as an integrated system, and it has been shown that for fulfilling service quality improvement demands, more number of wireless services must be optimized for more than one requirement. In regard to the ISTCNs, CR with NOMA has been proposed, which shows high efficiency due to complete access of the spectrum. Considering a UIoT network, the space-air-ground area can be covered and a hybrid blockchain technique can be implemented. Such a network shows high security and reliability simultaneously enabling massive data interconnection and data authenticity. Further, for UDNs it is required to provision the users with safety. Such networks have been shown to minimize the user access delay and improve the network throughput.

Overall, to implement SS in B5G wireless networks, the research community has already presented solutions in the form of O-RAN, which enhances the choice of users and the interoperability of different vendors and protocol stacks. The geographical coverage will be increased by integrating the terrestrial with the non-terrestrial network, and the SE will be enhanced via SS between the systems. Further, ML techniques will find advanced applications within the SS-enabled systems, specifically in regard to spectrum allocation and management. Table VIII summarizes all the recent related studies that have implemented SS in the B5G intelligent wireless networks.

### *C. Challenges*

The most significant challenges for implementing SS in B5G intelligent wireless networks are summarized next.

1) In B5G wireless networks, there will be an increase in the complexity of operations as there will be more spectrum access options, and the spectrum fragmentation techniques will also vary according to the country.
2) Advanced CA methods will be required when operating over different spectrum bands, which range between licensed spectrum to unlicensed and shared spectrum.
3) The dynamic nature of spectrum management will require an AI enabled intelligent design of the wireless network architecture. Further, in view of effectiveness of the spectrum management policies, it must be ensured that services are rolled-out and obstacles to entry are removed such that there occurs innovation in a globalising society with increasing demand for spectrum frequencies.
4) In regard to unlicensed access, the traditional LBT method will not be effective in the mm-waves and THz bands. This is due to the fact that in higher spectrum bands, transmitted beams will not be easily detectable owing to the carrier sense directivity and this will result in inaccurate spectrum occupancy decisions. Hence, there is a need to advance the LBT schemes in accordance to the requirements of B5G networks.
5) Multi-connectivity integrating multiple radio access technologies with the traffic flow aggregation among various bands must be addressed.
6) A major challenge will be to define SS rules, conditions, and mechanisms for ensuring that primary users do not encounter interference and that the new users obtain favorable deployment opportunities.
7) Developing explainable, actionable, and efficient ML techniques for SS remains an open research challenge. Also, advanced SS methods are required for implementation in the ISTCNs.
8) While addressing the multiple significant challenges in B5G wireless networks, it will be required to account for the appropriate trade-offs such as between EE and SE.

## V. Future Research Directions

In this section, we summarize key learnings from detailed discussions throughout the paper on SS techniques in 5G and B5G wireless networks. We also present a summary of the SS initiatives for intelligent wireless networks by various standardization/regulatory bodies. Finally, we detail the multiple research directions to solve existing spectrum-related issues in intelligent wireless networks.

### A. Lessons Learnt

It is clear that SS is an efficient method for addressing the spectrum scarcity issues, which will be challenging in coming intelligent wireless networks. Such networks will need efficient SS techniques, which can be supported by multiple advanced technologies. The CR technology is one such candidate for implementing SS as it can handle fluctuations of the available spectrum range and varied demands for the QoS. Further, CR can provide solutions when various spectrum allocation policies are implemented. An alternate solution is the use of ML techniques, which will leverage learning within the wireless networks, in turn addressing issues of complex interactions between the co-existing networks. To enhance usage of spectrum and alleviate shortage of spectrum, the existing centralized architecture will not be efficient since it is non-transparent and highly vulnerable to multiple attacks. Blockchain-based dynamic SS is a good candidate solution in such a case as it enables decentralization, openness, transparency, immutability, and auditability[153].

The SS techniques implemented in the 5G wireless networks cannot be readily extended to B5G wireless networks. Indeed, the latter will require novel SS techniques, and only a few studies have contributed in this regard. Among the formulated and implemented SS techniques for B5G there is the O-RAN solution to enhance the user(s) selection and interoperability of multiple vendors and protocol stacks. Also, integrating terrestrial network with the non-terrestrial network is presented as a solution to increase the geographical coverage. Hence, it will be possible to increase the SE between these systems. Further, ML techniques have been implemented for advance applications within the SS-enabled systems to provide spectrum allocation and management solutions.

### B. Standardization

Various regulatory bodies have been involved in standardizing the SS techniques. The Licensed Spectrum Access framework targets the 2.3 GHz band as specified by the Radio Spectrum Policy Group (RSPG) [154], Working Group Frequency Management (WGFM) entrusted Project Team FM53 [152] and Project Team FM52 [155], [156]. In [157], ETSI has specified various spectrum access methods for meeting requirements of multiple verticals. The SS techniques in licensed bands can provide local area services with high QoS. Further, ETSI regulation has also harmonized specifications for 5GHz and 60GHz bands [158], [159]. The issues related to Dynamic Spectrum Sharing have been covered in 3GPP Rel-15, which advanced in Rel-16 with an aim for guaranteeing coexistence between LTE and NR systems [160], [161], [162], [163]. The specifications of LTE-Unlicensed technology, including Carrier Sense Adaptive Transmission (CSAT) procedure, have been determined by LTE-U Forum [164], [165]. In [166], ITU has introduced and studied a few key factors that significantly influence the valuation of the frequency spectrum. The Telecom Regulatory Authority of India (Trai) will be issuing a consultation article to ease the rules for SS between the MNOs [167]. The aim will be to ensure more efficient utilization of scarce natural resource while simultaneously offering additional monetising opportunities to the carriers. This will allow MNOs to leverage evolving technologies, monetise airwaves, and discourage idle airwaves. In [168], after summarizing spectrum allocation for the public commercial operators, it is stated that the 700 MHz band may also be reused for LTE by various operators in Finland. It is further cleared that this will also be the case for 2 GHz and 2.6 GHz FDD bands which may also be leveraged for 5G.

TABLE IX
SUMMARY OF POTENTIAL RESEARCH CHALLENGES AND FUTURE SCOPE IN SS WITH POSSIBLE DIRECTIONS TO OBTAIN THE SOLUTIONS.

| S. No. | Existing Challenge | Proposed Direction |
|---|---|---|
| 1. | Fading environment where the channel statistics change over time, and data transmission is hardly affected in the presence of a Doppler shift. | ITU channel models are considered to examine the suitability of different channel estimation techniques for high mobility environments in the presence of deep fading and high Doppler shifts, where a minimal number of random pilots are inserted in each OFDM symbol equally spaced locations.[130] |
| 2. | (i) Understanding of sources of data used to develop Link Quality Estimation (LQE) models, (ii) applicability of LQE models to HetNets to incorporate multi-technology nodes, and (iii) broader and deeper understanding of link quality in multiple scenarios [131]. | To conduct synthesis of artificial data using GAN networks [132]. This is challenging since conducting such synthesis could potentially introduce unwanted bias to existing data, even though for specific applications numerous suitable examples of this method can be found in literature, such as wireless channel modeling [133], [134]. |
| 3. | Challenge in THz and mm-wave communication is severe path and absorption loss. The mm-Wave and THz carriers cannot penetrate physical obstacles making them highly susceptible to blockage. [135]. | Beamforming increases transmission range. The prediction techniques can be designed to infer the incoming blocakge in beam direction. Thus, by employing DL techniques it is possible to 'steer' beam in another direction in real time. [136]. |
| 4. | The B5G devices require higher energy compared to past generations as they operate in higher bands. Hence, power consumption and EE are key challenges to be considered [137]. | Level of super EE and even battery-free IoE devices, e.g., energy harvesting in building automation and smart homes must be investigated. |
| 5. | Intelligent Proactive Caching and Mobile Edge Computing- Based on popularity/demand rate to buffer data at nodes (IoT devices, BSs, etc.) level intelligently. | ML advised enhanced caching. [138]. |
| 6. | The development and management of micro licensing model with detailed micro licensing elements such as license areas, time durations, needed obligations, and the protection criteria for both (co-channel and adjacent channel) channel usage between micro licenses and possible incumbents. The existing interference techniques for coordination by considering ideas from individual and general authorization regimes , change interference conditions to allow dynamic adaptation and different protection levels of interference. | To use blockchain technology to ensure transparency in transactions spectrum which could become many with large number of local licenses with shorter time durations that will need an efficient marketplace [139]. |
| 7. | The limited data amount from 5G spectrum assignment decisions and high uncertainty of market data for spectrum study. [113]. | More detailed analysis will provide insights into potential value of spectrum options for different vertical domains served by 5G networks by using empirical data for local licensing in 3.6 GHz band. |
| 8. | Interference coordination problem [140]. | Communication quality is mainly affected by neighbor users in D2D communication. Hence, analysis of neighbor users' information will be conducted to conclude the interference problem. |
| 9. | Blockchain, scalability and performance issues [98] | Methodologies to improve the throughput |
| 10. | Storage and computation burden on resource constrained IoT devices [141] | The capacity of blockchain will be large if transaction data are stored in the chain. Hence, it needs to maintain the chain in such cases. |
| 11. | Blockchain security and privacy [142] | The big data resources need High-security guarantees and this can be provided with decentralized management, as it is associated with reliability and authentication of Blockchain. |
| 12. | To determine the omnidirectional radiation characteristics of a mobile phone device by measuring the Effective Isotropic Radiated Power (EIRP) for a specific beam ID [143]. | The EIRP with total distribution function enables evaluation of spherical coverage of a 5G user equipment [143]. |
| 13. | Advancement on ML and related methods are required for assisting self-tuning surfaces in introducing autonomous operations.[144] | Accurate positioning considering radio dynamics and electromagnetically tuneable surfaces. |
| 14. | It is important to be data driven and leverage AI methods for managing 5G networks efficiently [145]. | Study on 3GPP-17 investigated high-level principles, functional framework, potential use cases, and associated solutions for AI-enabled RAN intelligence [146]. |
| 15. | Remote control and factory automation for industries [147]. | A study on highly efficient actualization method of a network (individual network) specialized for industry that is different from the best effort type service of a public network is in progress. |
| 16. | Adapting network resources for meeting end-user's demands of QoS simultaneously reducing consumption of network power [148]. | M-MIMO and lean carrier designs must be investigated |
| 17. | Change in the ecosystem | Different stakeholder roles provide physical infrastructure, equipment, and data, under regulatory framework set by policymakers [149]. |
| 18. | Convergence guarantees when devices may differ in terms of hardware, network connection, and battery power. | Asynchronous Federated Learning model for heterogeneous network [150]. |
| 19. | Improving Dynamic SE [151] | Adding extended dimensions of observation, such as the temporal and spatial statistics of the spectra. |
| 20. | Licensed Spectrum Access (LSA) framework minimizes investment in extra-base stations. For this case, static interference management is based on predefined information in the Licensed Spectrum Access Repository. There also exists highly dependent nature of LSA framework on incumbents' partnership [152]. | Address the time required for proper evacuation of LSA licensees from band, and evaluate consequences when other incumbents in band are not willing for sharing. |

### C. Summary

To summarize, it is clear from our detailed discussions that factors such as SE gains, protection from interference, network congestion, and mobility support must be accounted for when selecting any SS technique. Also, no 'best' SS technique exists since it largely depends on the application for which it is intended. Hence, implementing advance SS techniques in B5G wireless networks requires significant research efforts. In Table IX, we list the key existing challenges and propose relevant research directions aiming to address these issues.

## VI. CONCLUSION

In this article, we reviewed various SS techniques for intelligent wireless networks. The multiple spectrum utilization opportunities arsing from the existing SS methods were highlighted. Next, the various types of SS techniques were detailed. Moreover, the main limitations of the existing SS methods were highlighted, and relevant solutions offered by the B5G intelligent wireless networks were presented. The focus, contributions, and significant features of each SS technique were discussed, and relevant technologies in regard to SS and multiple applications related to the 5G/B5G technology were also reviewed. Lastly, future research directions in response to the multiple issues and challenges requiring timely solutions were presented.

This survey article has evinced that due to significant advancements in wireless networks, the existing SS techniques that are applicable to the previous or existing wireless network generations will not be applicable to the B5G wireless networks. Also, the centralized SS techniques will be inefficient, and blockchain-enabled decentralized methods will be required. Further, much research efforts are required to overcome the concerns related to a combination of various spectrum bands operating with heterogeneous features. Finally, key challenges will include defining the SS rules, conditions, and mechanisms to ensure that the incumbent users face the least amount of interference and the new users obtain favorable deployment opportunities.

Lastly, this comprehensive survey will extend research on important issues related to SS techniques in the B5G wireless networks and will aid in successfully implementing the multiple B5G technology-enabled applications.

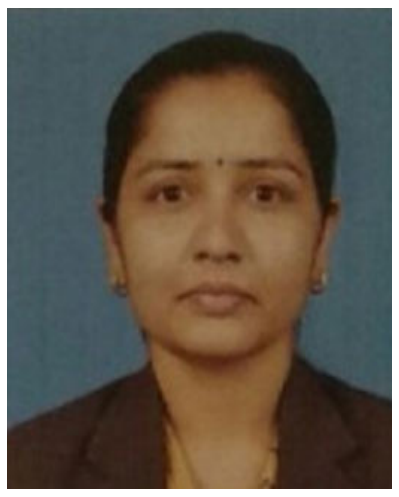

**Anita Patil** received Bachelor of Electronics and Communication Degree from Visvesvaraya Technological University, India, and Masters degree in Digital Electronics Visvesvaraya Technological University India, in 2014. Currently, she is pursuing the Ph.D. degree from Visvesvaraya Technological University India. She serves as an Assistant Professor in the Department of Electronics and Communication Engineering, S.G. Balekundri Institute of Technology, India. Her research interests include 6G communications, AI and ML.

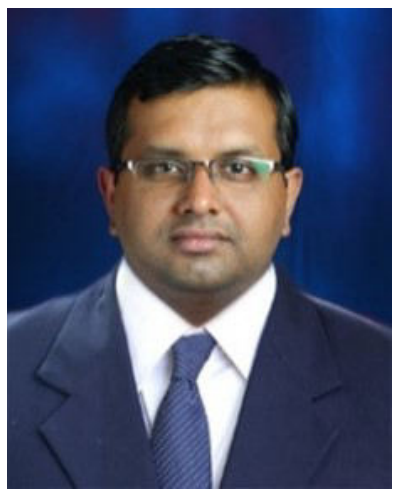

**Sridhar Iyer** received the M.S. degree in Electrical Engineering from New Mexico State University, U.S.A in 2008, and the Ph.D. degree from Delhi University, India in 2017. He received the young scientist award from the DST/SERB, Govt. of India in 2013, and Young Researcher Award from Institute of Scholars in 2021. He is the Recipient of the 'Protsahan Award' from IEEE ComSoc, Bangalore Chapter as a recognition to his contributions towards paper published/tutorial offered in recognized conferences/journals (during Jan 2020 - Sep 2021). He has completed two funded research projects as the Principal Investigator, and is currently involved in an on-going funded research projects as the Principal Investigator. His current research focus includes semantic communications and spectrum enhancement techniques for Intelligent wireless networks, and efficient design and resource optimization of the elastic optical networks enabled by space division multiplexing. He has published over 100 reviewed articles in the aforementioned areas. Currently, he serves as an Associate Professor in the Dept. of ECE, KLE Technological University, Dr MSSCET, Belagavi, Karnataka, India. Google Scholar: https://scholar.google.co.in/citations?user=2hbORHEAAAAJ&hl=en

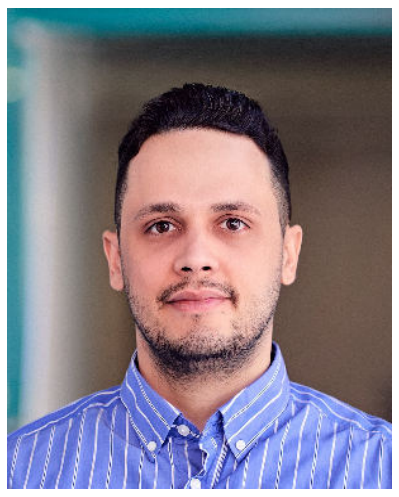

**Onel López** received the B.Sc. (1st class honors, 2013), M.Sc. (2017) and D.Sc. (with distinction, 2020) degree in Electrical Engineering from the Central University of Las Villas (Cuba), the Federal University of Paraná (Brazil) and the University of Oulu (Finland), respectively. From 2013-2015, he served as a specialist in telematics at the Cuban telecommunications company (ETECSA). He is a collaborator to the 2016 Research Award given by the Cuban Academy of Sciences, a co-recipient of the 2019 IEEE European Conference on Networks and Communications (EuCNC) Best Student Paper Award, the recipient of both the 2020 best doctoral thesis award granted by Academic Engineers and Architects in Finland TEK and Tekniska Föreningen i Finland TFiF in 2021 and the 2022 Young Researcher Award in the field of technology in Finland. He is co-author of the book entitled "Wireless RF Energy Transfer in the massive IoT era: towards sustainable zero-energy networks", Wiley, Dec 2021. He currently holds an Assistant Professorship (tenure track) in sustainable wireless communications engineering in the Centre for Wireless Communications (CWC), Oulu, Finland. For more info: https://sites.google.com/view/onellopez

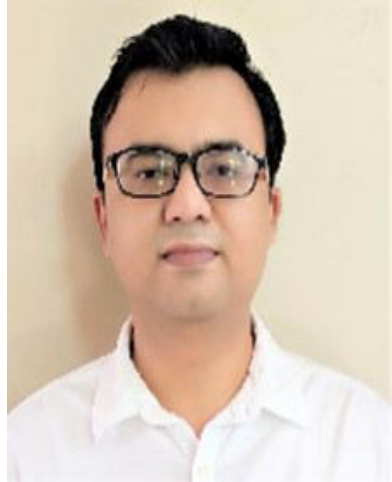

**Rahul Jashvantbhai Pandya** completed his M.Tech. from the Electrical Engineering Department, Indian Institute of Technology, Delhi, New Delhi, in 2010. He completed his Ph.D. from Bharti School of Telecommunication, Indian Institute of Technology, Delhi, in 2014. He worked as the Sr. Network Design Engineer in the Optical Networking Industry, Infinera Pvt. Ltd., Bangalore, from 2014 to 2018. Later, from 2018 to 2020, he worked as the Assistant Professor, ECE Department, National Institute of Technology, Warangal. Currently, he is working in Electrical Engineering department at Indian Institute of Technology, Dharwad. His research areas are Wireless Communication, Optical Communication, Optical Networks, Computer Networks, Machine Learning, and Artificial Intelligence. He is working on multiple projects, such as SERB, SPARC, SGNF, and RSM. Google Scholar: https://sites.google.com/iitdh.ac.in/rpandya

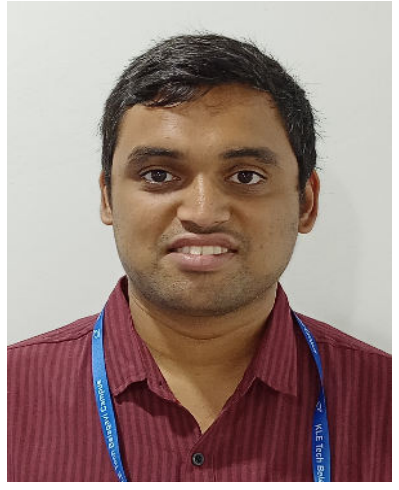

**Krishna Pai** received the Bachelor's degree in Electrical and Electronics Engineering from KLE Dr. M.S. Sheshgiri College of Engineering and Technology, Belagavi, India in 2021. Currently, he serves as a Teaching Assistant in the Department of Electronics and Communication Engineering, KLE Technological University's Dr. M.S. Sheshgiri College of Engineering and Technology, Belagavi, India. His research interests include Artificial Intelligence, Machine Learning, Internet of Things, Electrical Vehicles, Battery Management systems, Embedded Systems and Communication Theory. Presently, he is involved in research over numerous funded projects and publication of the related articles.

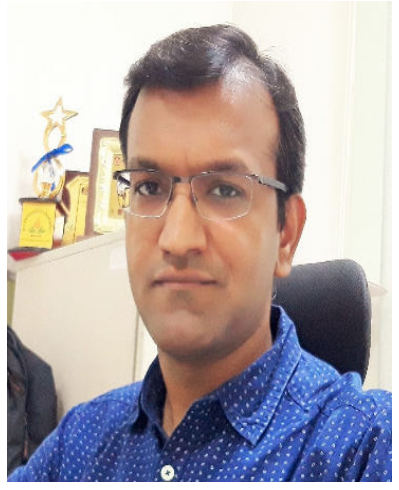

**Anshuman Kalla (Senior Member IEEE)** is working as Professor at the Department of Computer Engineering, CGPIT, Uka Tarsadia University (UTU), India. Dr. Kalla has more than twelve years of teaching and research experience. Dr. Kalla is a senior IEEE member. He has worked as a Postdoctoral Visiting Researcher at Center for Wireless Communications (CWC), University of Oulu, Finland. He graduated as an Engineer from Govt. Engineering College Bikaner in 2004. He did Master of Science in Telecommunications and Wireless Networking from ISEP, Paris, France in 2008 and another Master from University of Nice Sophia Antipolis, France in 2011. He obtained a Ph.D. degree in 2017. Dr. Kalla was recipient of Master's scholarships for pursuing both the Master programs. Dr. Kalla has delivered invited sessions and talks during various Short Term Courses, Faculty Development Programs, Workshops and Conferences. In particular, he has delivered tutorials at IEEE 5G World Forum, 2021, IEEE ANTS 2020, and IEEE 5G World Forum, 2020. He has published papers in reputed international journals such as Elsevier (JII, IPM, JNCA, COMNET, ICT Express), IEEE Consumer Electronics Magazine, IEEE ComSoc, IEEE Computer, and IEEE EMR. His area of interests are Blockchain, 5G, 6G, IoT, Information Centric Networking, Software Defined Networking, Next Generation Networks. For more info: https://sites.google.com/site/kallanshuman

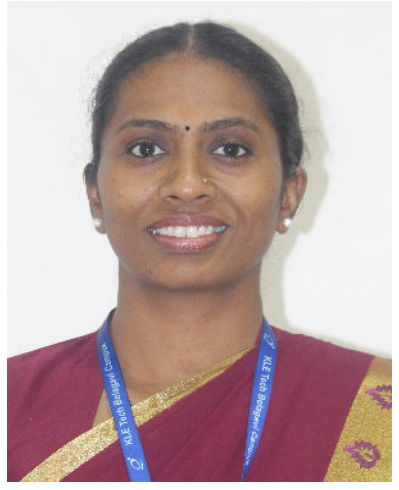

**Rakhee Kallimani** received the M.Tech degree in Embedded Systems from Calicut University, in 2007 and B.E. degree from B V Bhoomraddi College of Engineering and Technology affiliated to Visvesvaraya Technological University in 2003 in Electrical and Electronics. She defended the Ph.D. studies at Visvesvaraya Technological University in November 2022. She has previously served at BVBCET as Lecturer from 2003-2009. Currently, she serves as an Assistant Professor at KLE Technological Univerity Dr MSSCET, Belagavi. Her area of Interest is Embedded, Systems, Wireless Sensor Networks and Electrical Vehicles.